\documentclass[a4paper,footinbib,nofootinbib]{revtex4-1}
\usepackage{graphicx,color,amsmath}
\usepackage{hyperref}

\newcommand{\eps} {\epsilon}
\def\met        {E\!\!\!\!/_T}
\newcommand{\beq} {\begin{equation}}
\newcommand{\eeq} {\end{equation}}

\begin{document} 

\title{Flavor violating signatures of lighter and heavier Higgs bosons within the Two Higgs Doublet Model Type-III at the LHeC}

\date{\today}

\author{S.P. Das}
\affiliation{ Instituto de F\'{\i}sica, Benem\'erita Universidad Aut\'onoma de Puebla, \\ Apdo. Postal J-48, C.P. 72570 Puebla, Puebla, M\'exico and Facultad de Ciencias F\'isico-Matem\'aticas,
Benem\'erita Universidad Aut\'onoma de Puebla,
\\ Apdo. Postal 1364, C.P. 72570 Puebla, Puebla, M\'exico. }
\email{sprasad@ifuap.buap.mx}

\author{J. Hern\'andez-S\'anchez}
\affiliation{Facultad de Ciencias de la Electr\'onica,
  Benem\'erita Universidad Aut\'onoma de Puebla,
\\ Apdo. Postal 542, C.P. 72570 Puebla, Puebla, M\'exico \\ 
  and Dual C-P Institute of High Energy Physics, M\'exico.}
\email{jaime.hernandez@correo.buap.mx}

\author{S. Moretti}
\affiliation{School of Physics and Astronomy, University of Southampton, Highfield, Southampton SO17 1BJ, United Kingdom, and Particle Physics Department, Rutherford Appleton Laboratory, Chilton, Didcot, Oxon OX11 0QX, United Kingdom}
\email{s.moretti@soton.ac.uk}

\author{A. Rosado,}
\affiliation{ Instituto de F\'{\i}sica, Benem\'erita Universidad Aut\'onoma de Puebla, \\ Apdo. Postal J-48, C.P. 72570 Puebla, Puebla, M\'exico.}
\email{rosado@ifuap.buap.mx}

\author{R. Xoxocotzi}
\affiliation{Facultad de Ciencias F\'isico-Matem\'aticas,
Benem\'erita Universidad Aut\'onoma de Puebla,
\\ Apdo. Postal 1364, C.P. 72570 Puebla, Puebla, M\'exico.}
\email{xoxo\_reyna@yahoo.com.mx}

\begin{abstract}
 We analyze the prospects for observing the light and heavy CP-even neutral 
  Higgs bosons ($\phi= h$ and $H$) in their decays into flavor violating $b \bar s$ channels
  (including charge conjugation) at the proposed Large Hadron electron Collider (LHeC), with $\sqrt s \approx 1.3$ TeV, 
in the framework of a 2-Higgs Doublet 
  Model (2HDM) Type-III, assuming a four-zero texture in the Yukawa matrices and a general Higgs 
  potential. We consider theoretically consistent scenarios in  agreement with  current experimental data from
  flavor and Higgs physics. We investigate the charged current production 
  process $\nu_e \phi q$ in presence of  flavor violating decays of the Higgs bosons, that 
  lead to a 3-jets + $\met$ signature. We demand exactly two jets, one tagged $b$-jet and one light-flavor jet, all 
  in the central rapidity region. The remaining jet (originated by the remnant quark $q$) is tagged 
  in the forward or  backward regions and this together with a central jet veto (not more than one light-flavor jet) 
  are essential criterions to enhance the signal--to--background rates. We consider the most 
  relevant Standard Model (SM) backgrounds, treating $c$-jets 
  separately from light-flavor and gluon ones, while allowing for mis-tagging. We find that the 
  SM-like Higgs boson, $h$, would be accessible within  several parameter configurations of our model at  approximately the
  1-2$\sigma$ level with 100 fb$^{-1}$ of data. We also find that the heaviest neutral 
  Higgs boson, $H$, with mass up to 150 GeV, would have a 1$\sigma$ significance for the same data sample. At the end of  the LHeC 
  running, one would have ten times data accumulation and for all the Higgs masses the 
  significances are enhanced so as to allow for detection of both the $h$ and $H$ state. 
Hence, one of 
  the most viable extensions of 2HDMs with Flavor Changing Neutral Currents (FCNCs) generated 
  at tree level but controlled by a four-zero texture approach in the Yukawa matrices, as opposed to 
the adoption of ad hoc discrete symmetries, could be put under scrutiny at a future $ep$ machine.
\end{abstract}


\maketitle
\flushbottom

\section{Introduction}
\label{sec:intro}

The Standard Model (SM) is well established by now after the discovery of a Higgs boson by the ATLAS \cite{atlash} 
and CMS \cite{cmsh} experiments at the Large Hadron Collider (LHC). However, when one considers some theoretical aspects of the SM, for example, lepton number violation, which is already manifested  in the form of small 
but non-zero neutrino masses and actively searched for in the other two fermionic sectors 
(e.g., in $\mu \to e \gamma$ decays by the MEG experiment \cite{meg} and 
in $B$-physics by BaBar \cite{babar} and Belle \cite{belle} \footnote{In fact, also  
top-quark flavor violating decays into charm quarks and Higgs bosons are currently under investigation at the LHC.},
one necessarily has to postulate New Physics (NP) Beyond the SM (BSM). 
By combining the evidence of Higgs states and the presence in Nature of flavor violation, then one is well  motivated in searching for evidence of BSM physics in a context where the two aspects merge, i.e., in flavor violating Higgs boson decays.
In general, limited to the Higgs sector, several BSM scenarios
have been invoked by introducing extra Higgs singlets, doublets and/or triplets. As the Higgs boson
discovery at the LHC is consistent with a doublet structure, we refrain here from considering BSM constructions with either of
the other two aforementioned Higgs representations. Therefore, in staying with multi-Higgs doublet structures, the simplest of such Higgs scenarios is the so called 2-Higgs Doublet Model (2HDM) \cite{Branco:2011iw,Gunion:2002zf}, which will be the theoretical focus of our study.

Among the many phenomenological sides of a 2HDM, we are indeed 
concerned here  with flavor violating Higgs boson decays
in the quark sector, building on previous works of ours, see, e.g., Refs. \cite{Cordero-Cid:2013sxa, Felix-Beltran:2013tra, HernandezSanchez:2012eg,HernandezSanchez:2013xj,
HernandezSanchez:2011fq,
Akeroyd:2014cma, DiazCruz:2004pj,DiazCruz:2004tr}. However, plentiful of studies, also including lepton
flavor violating scenarios, exist, some specific to 2HDMs and others adaptable to their case: see for an
incomplete list, e.g., Refs.
\cite{DiazCruz:1999xe,higgslfv,
Sierra:2014nqa,Kopp:2014rva,Greljo:2014dka,
Dery:2014kxa,Vicente:2014mya,Boucenna:2015zwa,GomezBock:2005hc,deLima:2015pqa,
Dorsner:2015mja,Omura:2015nja,cvetic,Fernandez:2009vr,cotti,BarShalom:2008fq,Larios:2004mx}.

The actual search scope for Higgs bosons in flavor violating modes at the LHC has also been actively studied, see  \cite{Assamagan:2002kf,Han:2000jz,Craig:2012vn,Craig:2013hca,
Dumont:2014wha,Brownson:2013lka,Coleppa:2013dya,Holdom:2014boa}\footnote{Current experimental results at the LHC include both 
ATLAS \cite{atlasconf201327}  and CMS \cite{Khachatryan:2015kon} analyses.}. It has also been investigated in the context of
a future $e^+ e^-$  \cite{Johansen:2015nxa} and $\gamma\gamma$ collider \cite{HernandezSanchez:2011fq}. 
Prospects at a future hadron collider have been investigated in 
\cite{Brownson:2013lka}. 

Herein, we are particularly motivated by a possible enhancement of  flavor violating quark decays 
($\phi \to b \bar s$) of  intermediate mass Higgs bosons (below the top-quark mass) 
and we will focus on the possibility to access such signatures at the possibly upcoming Large Hadron electron Collider (LHeC).
The LHeC facility \cite{cern:lhec} presently discussed as possible at CERN in the near future 
would be a Deep Inelastic Scattering (DIS) experiment at the TeV scale, 
with center-of-mass energy of around 1.3 TeV. In comparison to the another 
recently closed  (in 2007) DIS experiment (the Hadron-Electron 
Ring Accelerator (HERA) \cite{heraphys} at DESY of around 320 GeV in energy with integrated luminosity 
of around 0.5 fb$^{-1}$), the LHeC might deliver data samples of approximately 
100 fb$^{-1}$ and with a higher detector coverage. Further, the overall kinematic range (in $x$ and $Q^2$)
accessible at LHeC is 20 times larger than at HERA.

While the primary task of  a collider like the LHeC will be in-depth studies of QCD, the machine also affords some scope
to study Higgs bosons decaying via flavor violating processes
\cite{Sarmiento-Alvarado:2014eha}.  Our objective in this paper is to study the 
feasibility of finding two CP-even neutral Higgs bosons, the SM-like Higgs state $h$ and its heavier
counterpart $H$,  at the upcoming LHeC assuming as BSM framework a 2HDM
 Type-III (henceforth 2HDM-III for short), which embodies a four-zero  texture  approach in the  Yukawa matrices as the mechanism to control 
Flavor Changing Neutral Currents (FCNCs). As we have shown in previous analyses, this is precisely the framework
which establishes $\phi\to b\bar s$ + c.c. as an hallmark signature of quark flavor violation in the Higgs sector whose
detectability is under investigation at the LHC and future $e^+e^-$ and $\gamma\gamma$ collider. 

The plan of our paper is as follows. In the next Section we describe briefly the theoretical structure of the 2HDM-III with a four-zero texture embedded in the Yukawa matrices.
In Section III, we demarcate the allowed 2HDM-III parameter space in presence of both theoretical 
and experimental constraints. 
In Section IV, we explain the characteristics of the Higgs boson signal from  
charged current production. We introduce the most important SM backgrounds and finally, we carry out both a parameter
space scan and a   signal-to-backgrounds analysis for some characteristic benchmarks by 
adopting a simple cut-based optimization to isolate $\phi\to b\bar s$ + c.c. events.
In section V, we recap and present our conclusions.

\section{The Higgs-Yukawa sector of  the 2HDM Type-III}
\label{sec:thdma}

In the 2HDM, the two Higgs scalar doublets, 
$\Phi^\dag_1=(\phi^-_1,\phi_1^{0*})$ and $\Phi^\dag_2=(\phi^-_2,\phi_2^{0*})$,  have the same hypercharge $+1$
such that both couple to the same quark flavor. Since a specific four-zero texture is implemented as a flavor symmetry in the Yukawa sector, this is the mechanism which controls  FCNCs so that discrete symmetries in the Higgs potential are not needed
\cite{Cordero-Cid:2013sxa, Felix-Beltran:2013tra,HernandezSanchez:2012eg,
HernandezSanchez:2013xj,HernandezSanchez:2011fq}.  
Then the most general $SU(2)_L \times U(1)_Y $ invariant scalar potential, 
following ~\cite{Gunion:2002zf}, can be written as:
\begin{eqnarray}
V(\Phi_1,\Phi_2)&=&\mu^2_1(\Phi_1^\dag
\Phi_1)+\mu^2_2(\Phi^\dag_2\Phi_2)-\left(\mu^2_{12}(\Phi^\dag_1\Phi_2)+{\rm
H.c.}\right) + \frac{1}{2}
\lambda_1(\Phi^\dag_1\Phi_1)^2 \\ \nonumber
&& +\frac{1}{2} \lambda_2(\Phi^\dag_2\Phi_2)^2+\lambda_3(\Phi_1^\dag
\Phi_1)(\Phi^\dag_2\, \Phi_2)
+\lambda_4(\Phi^\dag_1\Phi_2)(\Phi^\dag_2\Phi_1) \\ \nonumber
&& +
\left(\frac{1}{2} \lambda_5(\Phi^\dag_1\Phi_2)^2+\left(\lambda_6(\Phi_1^\dag
\Phi_1)+\lambda_7(\Phi^\dag_2\Phi_2)\right)(\Phi_1^\dag \Phi_2)+
{\rm H.c.}\right), \label{potential}
\end{eqnarray}
where all the parameters are assumed to be real\footnote{The  $\mu^2_{12}$, $\lambda_5$, 
$\lambda_6$ and $\lambda_7$ parameters could be complex in general, but for simplicity 
we assume these parameters to be  real.}, including the Vacuum Expectation Values (VEVs) of the 
scalar fields, hence there is no CP-violation. In general, by introducing 
some discrete symmetry $\Phi_1\to \Phi_1$ and $\Phi_2\to -\Phi_2$ , the scalar 
potential does not have the contributions of $\lambda_6$ and $\lambda_7$.

It has long been proposed that there are four possibilities to satisfy 
the Paschos-Glashow-Weinberg theorem \cite{glashow} in  2HDMs 
\cite{Branco:2011iw},\cite{Gunion:2002zf}. These are defined as  
follows: Type I (where one Higgs doublet couples to all fermions); Type-II (where one Higgs doublet couples to up-type 
quarks and the other to  down-type ones); Type-X (also called 
``Lepton-specific'', where the quark couplings are Type-I and the 
lepton ones Type-II); Type Y (also called ``Flipped'' model, 
where the quark couplings are Type II and the lepton ones Type-I).
With these two scalar doublets, there are eight fields but only five of 
them are physical (pseudo)scalar (``Higgs'') fields, which correspond to: two neutral
CP-even bosons $h$ (lighter)/$H$ (heavier),
one neutral CP-odd boson $A$ and two charged bosons $H^{\pm}$. 
The mixing angle $\alpha$ of the two neutral CP -even bosons $h$ and $H$ 
is another parameter of the 2HDM model. In total, the 2HDM model 
can be described by the parameters $\alpha$, $\beta$ (where $\tan\beta$ is 
the ratio of the VEVs of the two Higgs doublet fields) 
and the masses of the five Higgs particles. With these inputs one can 
estimate all the parameters that are present in the scalar potential, to be 
specific, the $\lambda$'s. These $\lambda$'s (together with various scalar 
mass parameters) enter the expressions of the theoretical constraints like: 
vacuum stability, unitarity, perturbativity and also various 
EW Precision Observables (EWPOs), for example, the oblique 
parameters. All these 2HDM types are fully compatible with the SM-like Higgs boson discovery. 

The flavor sector of  2HDMs is testable 
in low  as well as in high energy collider experiments. The tests have been 
carried out in the most general version of a 2HDM 
with a Yukawa four-zero texture, wherein the Yukawa couplings 
are proportional to the geometric mean of two fermions masses, 
$g_{ij} \propto \sqrt{(m_i m_j)} \chi_{ij}$ \cite{jhs1,jhs2}. 
As it was mentioned, a consequence  of this  is that the terms  of the scalar potential including $\lambda_6$ and $\lambda_7$ 
should now be taken into account.
This leads to  tri-linear and quartic self-couplings of the scalar fields 
\cite{Cordero-Cid:2013sxa,HernandezSanchez:2011fq} affecting the model phenomenology 
in one loop processes via  di-Higgs and tri-Higgs topologies, both in production and decay processes.
It has been shown that the EWPO $\rho$ can deviate from experimental bounds at one loop level, as long as the mass difference 
between charged Higgs bosons with CP-even/CP-odd masses is large, irrespective  of the value of $\lambda_6$ and $\lambda_7$. Hence, some level of
degeneracy between one neutral and the charged Higgs states is a precondition on the 2HDM spectra .  In our construction, 
the Yukawa Lagrangian \cite{Cordero-Cid:2013sxa} is given by:
{\small
\begin{equation}
 {\cal{L}}_{Y}  = -\Bigg(
Y^{u}_1\bar{Q}_L {\tilde \Phi_{1}} u_{R} +
                   Y^{u}_2 \bar{Q}_L {\tilde \Phi_{2}} u_{R} +
Y^{d}_1\bar{Q}_L \Phi_{1} d_{R}
 + Y^{d}_2 \bar{Q}_L\Phi_{2}d_{R} +Y^{{l}}_{1}\bar{L_{L}}\Phi_{1}l_{R} +Y^{{l}}_{2}\bar{L_{L}}\Phi_{2}l_{R} \Bigg),
\label{lag-f}
\end{equation} }
where $\Phi_{1,2}=(\phi^+_{1,2},
\phi^0_{1,2})^T$ refer to the two Higgs doublets, ${\tilde
\Phi_{1,2}}=i \sigma_{2}\Phi_{1,2}^* $. Besides,
the fermion mass matrices after EW  symmetry breaking are given,
from 
Eq. (\ref{lag-f}), by: $ M_f= \frac{1}{\sqrt{2}}(v_{1}Y_{1}^{f}+v_{2}Y_{2}^{f})$, $f = u$, $d$, $l$,
assuming that both Yukawa matrices $Y^f_1$ and $Y^f_2$ have the
four-zero-texture form and are Hermitian \cite{HernandezSanchez:2012eg,jhs1,DiazCruz:2004pj}. 
After diagonalisation,
$ \bar{M}_f = V_{fL}^{\dagger}M_{f}V_{fR}$, one has
$\bar{M}_f=\frac{1}{\sqrt{2}}(v_{1}\tilde{Y}_{1}^{f}+v_{2}
\tilde{Y}_{2}^{f})$,
where $\tilde{Y}_{i}^{f}=V_{fL}^{\dagger}Y_{i}^{f}V_{fR}$.
One can obtain a compact and generic form for 
 the rotated matrix $\tilde {Y}^f_n$\footnote{We have shown in several parametrisations that this structure corresponds, as a particular case,  to the Cheng and Sher ansatz \cite{jhs1,DiazCruz:2004tr,Felix-Beltran:2013tra,HernandezSanchez:2012eg}.}:
\begin{eqnarray}
\left[ \tilde{Y}_n^{f} \right]_{ij}
= \frac{\sqrt{m^f_i m^f_j}}{v} \, \left[\tilde{\chi}_{n}^f \right]_{ij}
=\frac{\sqrt{m^f_i m^f_j}}{v}\,\left[\chi_{n}^f \right]_{ij}  \, e^{i \vartheta^f_{ij}},
\label{cheng-sher}
\end{eqnarray}
\noindent
where the $\chi$'s are unknown dimensionless parameters of the model.
Following \cite{HernandezSanchez:2012eg}, one has a 
 generic expression for the couplings  of the Higgs bosons to the fermions  given as
{\small
\begin{eqnarray}
{\cal L}^{\bar{f}_i f_j \phi}  & = &
-\left\{\frac{\sqrt2}{v}\overline{u}_i
\left(m_{d_j} X_{ij} {P}_R+m_{u_i} Y_{ij} {P}_L\right)d_j \,H^+
+\frac{\sqrt2m_{{l}_j} }{v} Z_{ij}\overline{\nu_L^{}}{l}_R^{}H^+
+{H.c.}\right\} \nonumber \\
& &-
\frac{1}{v} \bigg\{ \bar{f}_i m_{f_i} h_{ij}^f  f_j h^0 + \bar{f}_i m_{f_i} H_{ij}^f  f_j H^0 - i \bar{f}_i m_{f_i} A_{ij}^f  f_j \gamma_5 A^0\bigg\},
\label{lagrangian-f}
\end{eqnarray}
where $\phi_{ij}^f$ ($\phi=h$, $H$, $A$), $X_{ij}$, $Y_{ij}$ and $Z_{i j}$ are defined as follows:
\begin{eqnarray}\label{hHA}
h_{ij}^d & = & \xi_h^d \delta_{ij} + \frac{(\xi_H^d-X \xi_h^d)}{\sqrt{2} f(X)} \sqrt{\frac{m_{d_{j}}}{m_{d_i}}} \tilde{\chi}_{ij}^d, \, \, \,  h_{ij}^\ell =  h_{ij}^d (d \to \ell, X\to Z),   \\
H_{ij}^d & = & \xi_H^d \delta_{ij} - \frac{(\xi_h^d+X \xi_H^d)}{\sqrt{2} f(X)} \sqrt{\frac{m_{d_{j}}}{m_{d_i}}} \tilde{\chi}_{ij}^d, \, \, \,  H_{ij}^\ell =  H_{ij}^d (d \to \ell, X\to Z),   \\
A_{ij}^d&=& -X \delta_{ij} + \frac{f(X)}{\sqrt{2}} \sqrt{\frac{m_{d_{j}}}{m_{d_i}}} \tilde{\chi}_{ij}^d,  \, \, \,  A_{ij}^\ell =  A_{ij}^d (d \to \ell, X\to Z),  A_{ij}^u =  A_{ij}^d (d \to u, X\to Y),\\
h_{ij}^u & = & \xi_h^u \delta_{ij} - \frac{(\xi_H^u+Y \xi_h^u)}{\sqrt{2} f(Y)} \sqrt{\frac{m_{u_{j}}}{m_{u_i}}} \tilde{\chi}_{ij}^u, \, \, \,     
H_{ij}^u =  \xi_H^u \delta_{ij} + \frac{(\xi_h^u-Y \xi_H^u)}{\sqrt{2} f(Y)} \sqrt{\frac{m_{u_{j}}}{m_{u_i}}} \tilde{\chi}_{ij}^u, \, \, \,   \\
X_{i j} & = &   \sum^3_{l=1}  (V_{\rm CKM})_{il} \bigg[ X \, \frac{m_{d_{l}}}{m_{d_j}} \, \delta_{lj}
-\frac{f(X)}{\sqrt{2} }  \,\sqrt{\frac{m_{d_l}}{ m_{d_j} }} \, \tilde{\chi}^d_{lj}  \bigg],
 \label{Xij} \\
Y_{i j} & = &  \sum^3_{l=1}  \bigg[ Y  \, \delta_{il}
  -\frac{f(Y)}{\sqrt{2} }  \,\sqrt{\frac{ m_{u_l}}{m_{u_i}} } \, \tilde{\chi}^u_{il}  \bigg]  (V_{\rm CKM})_{lj},
\label{Yij} \\
Z_{i j}^{l}& = &   \bigg[Z \, \frac{m_{{l}_{i}}}{m_{{l}_j}} \,
\delta_{ij} -\frac{f(Z)}{\sqrt{2} }  \,\sqrt{\frac{m_{{l}_i}}{m_{{l}_j}}  }
\, \tilde{\chi}^{l}_{ij}  \bigg],
\label{Zij}
\end{eqnarray} }
where $f(x) = \sqrt{1+x^2}$, $\xi_\phi^f$  are related to the trigonometric ratios (i.e., $\cos\alpha / \sin\beta$, $\sin\alpha / \sin\beta$, $\cos\alpha / \cos\beta$, $\sin\alpha / \cos\beta$)  and 
the parameters $X$, $Y$ and $Z$ can be related to $\tan \beta$ or $\cot \beta$, according to the various
 incarnations of 2HDMs  \cite{HernandezSanchez:2012eg} (see the Table \ref{tab:table1}). Taking into account that  the Higgs-fermion-fermion ($\phi f f$) coupling in the 2HDM-III is written as $g_{\rm 2HDM-III}^{\phi ff}= g_{\rm 2HDM-any}^{\phi ff} + \Delta g$, where $g_{\rm 2HDM-any}^{\phi ff}$ is the coupling $\phi f f$  in some of the 2HDMs with  discrete symmetry   and $\Delta g$ is the contribution of the four-zero texture\footnote{For example, one can recovers the Yukawa interactions given in 
Refs.~\cite{Grossman:1994jb, Akeroyd:2012yg,Aoki:2009ha} with $\chi_{ij}^f=0$.}, it was pointed out in \cite{HernandezSanchez:2012eg} that this Lagrangian could also 
represent a Multi-Higgs Doublet Model (MHDM) or an Aligned 2HDM (A2HDM) with additional 
flavor physics in the Yukawa matrices. 
\begin{table}[htbp]
  \centering
  \begin{tabular}{@{} |c|c|c|c|c|c|c|c|c|c| @{}}
\hline
\hline
    2HDM-III &$ X$ &$ Y$ & $Z$ &$ \xi_h^u$ & $\xi_h^d $& $\xi_h^\ell $ & $ \xi_H^u$ & $\xi_H^d $ & $\xi_H^\ell $\\ 
\hline
\hline
    2HDM-I-Like &$ -ct_\beta$ & $ct_ \beta$ &$ -ct_ \beta $& $c_\alpha / s_\beta $& $c_ \alpha / s_ \beta $ & $c_ \alpha / s_\beta $ &$ s_ \alpha / s_ \beta $ & $s_ \alpha / s_ \beta$ & $s_ \alpha / s_ \beta$ \\ 
\hline
     2HDM-II-Like  & $t_\beta$ & $ct_\beta$ & $t_\beta$ & $c_ \alpha / s_ \beta $ & $-s_ \alpha / c_ \beta $ & $-s_ \alpha / c_ \beta $ & $s_ \alpha / s_ \beta $ & $c_ \alpha / c_ \beta $ & $c_ \alpha / c_ \beta $ \\ 
\hline
     2HDM-X-Like  & $-ct_\beta$ & $ct_\beta$ & $t_\beta$ & $c_ \alpha / s_ \beta $ & $c_ \alpha / s_ \beta $ & $-s_ \alpha / c_ \beta $ & $s_ \alpha / s_ \beta $ & $s_ \alpha / s_ \beta $ & $c_ \alpha / c_ \beta $ \\ 
\hline
     2HDM-Y-Like  & $t_\beta$ & $ct_\beta$ & $-ct_\beta$ & $c_ \alpha / s_ \beta $ & $-s_ \alpha / c_ \beta $ & $c_ \alpha / s_ \beta $ & $s_ \alpha / s_ \beta $ & $c_ \alpha / c_ \beta $ & $s_ \alpha / s_ \beta $ \\ 
\hline
  \end{tabular}
  \caption{Parameters $\xi_\phi^f$ ($\phi= h$, $H$, $A$ with $f=u$, $d$, $\ell$ ),  $X, Y$ and $ Z$ as defined in Eqs. (\ref{hHA})--(\ref{Zij}) for four versions of the 2HDM-III. These parameters  are related to  the interactions of Higgs bosons with the fermions given in Eq. (\ref{lagrangian-f}). Here, $t_\beta =\tan \beta$, $ct_\beta= \cot \beta$, $s_\alpha= \sin \alpha$, $c_\alpha = \cos \alpha$, $s_\beta= \sin \beta$, $ c_\beta = \cos \beta$.}
  \label{tab:table1}
\end{table}

Here, we consider three different  incarnations of the 2HDM-III, which correspond to to the four 2HDM types already
described except the  lepton specific one, as here leptonic Branching Ratios (BRs) are dominant, whereas  we intend to look for an enhancement in the Higgs to $b\bar s$ decay because of  flavor violation.
We will finally show that, in different 
scenarios of the 2HDM-III, a substantial enhancement of the  BR($\phi \to b \bar s$) (including charge conjugation) is possible. We do so first via a parameter scan of the 2HDM-III at the inclusive level, followed by the detailed 
event generation analysis
of some benchmark scenarios amenable to phenomenological investigation.

\section{The 2HDM-III parameters and benchmarks}
\label{sec:thdmb}

In this section, we will perform a parameter scan of the 2HDM-III of interest from which we will extract our benchmark scenarios, all of which will be studied
in our final signal-to-background simulations, albeit we will show detailed results only for a subset of these for reasons of space.

First, we ought to explain the constraints we have enforced in our analysis. As for the experimental ones,
we have taken into account  recent experimental bounds 
from  flavor physics \cite{HernandezSanchez:2012eg,Felix-Beltran:2013tra}: i.e., from
$B \to \tau \nu_\tau$, $D \to \mu \nu$, $D_s \to  \ell \nu$, the semileptonic 
transition $ B \to D \tau \nu_{\tau}$, the inclusive decay $B\to X_s \gamma$, 
$B_0-B_0$ mixing, $B_s\to \mu^+ \mu^- $ and the radiative decay $Z \to b \bar{b}$.
(We have also imposed EWPO limits.) On the theoretical side, we have enforced
perturbativity, triviality, vacuum stability and unitarity constraints 
\cite{unitarity2hdm,Cordero-Cid:2013sxa}. In all the constraints mentioned above, the charged Higgs bosons masses are the  crucial parameter, 
as diagrams with $H^\pm$ states  co-exist alongside those involving 
the SM $W^\pm$ exchange diagrams. In this connection, alongside flavor and EWPO constraints, we have also accounted for those stemming from
Tevatron and  LHC searches 
\cite{HernandezSanchez:2012eg,HernandezSanchez:2013xj,Crivellin:2013wna,Deschamps:2009rh,Kanemura:2015mxa}\footnote{Current low energy constraints 
on the Higgs boson masses    have been studied very 
recently \cite{Biswas:2014uba,Das:2015qva}. }.
\subsection{Parameter scan}
\begin{figure}[ht!]
\centering
 \includegraphics[width=2.9in]{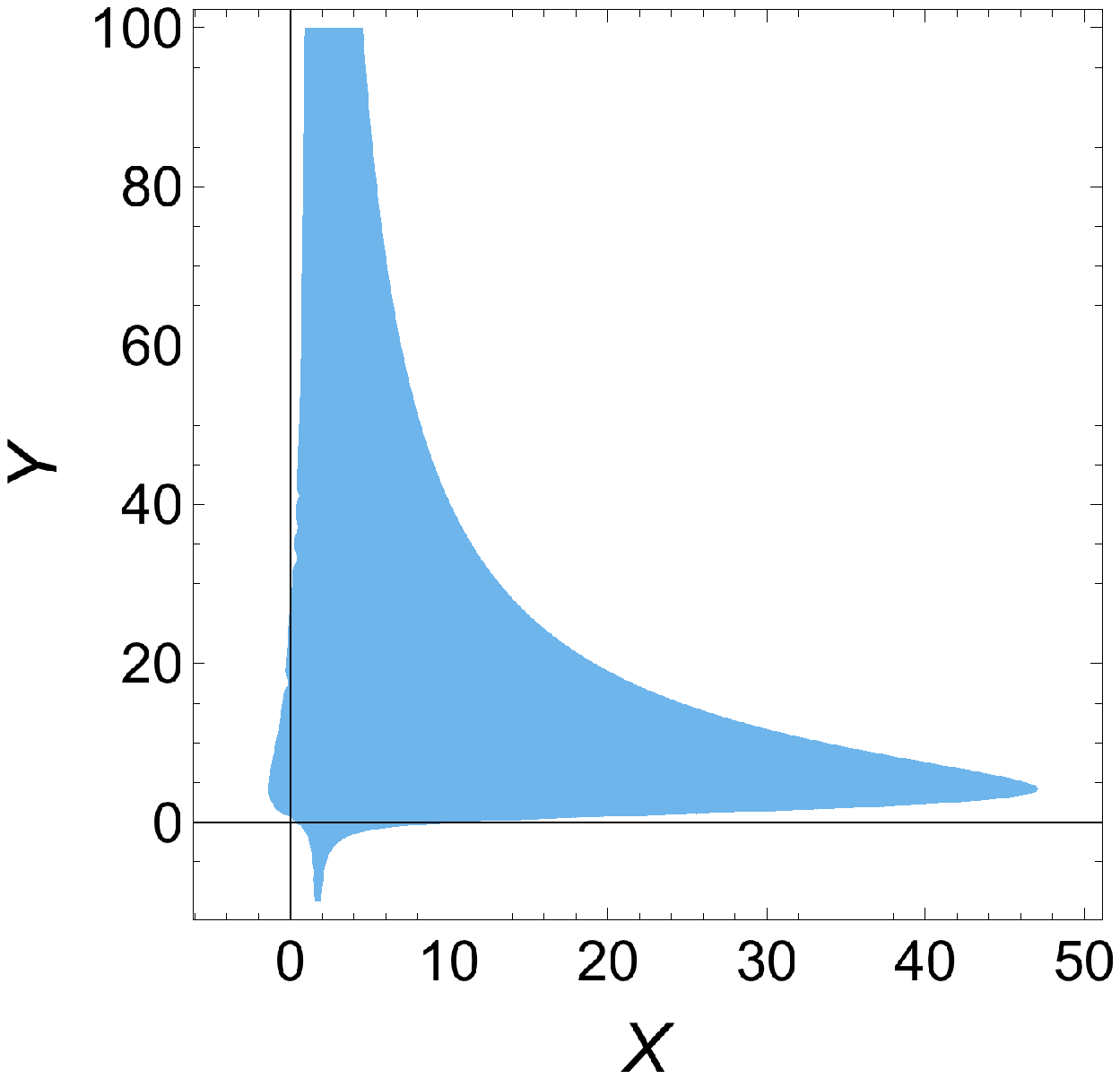}
 \includegraphics[width=2.9in]{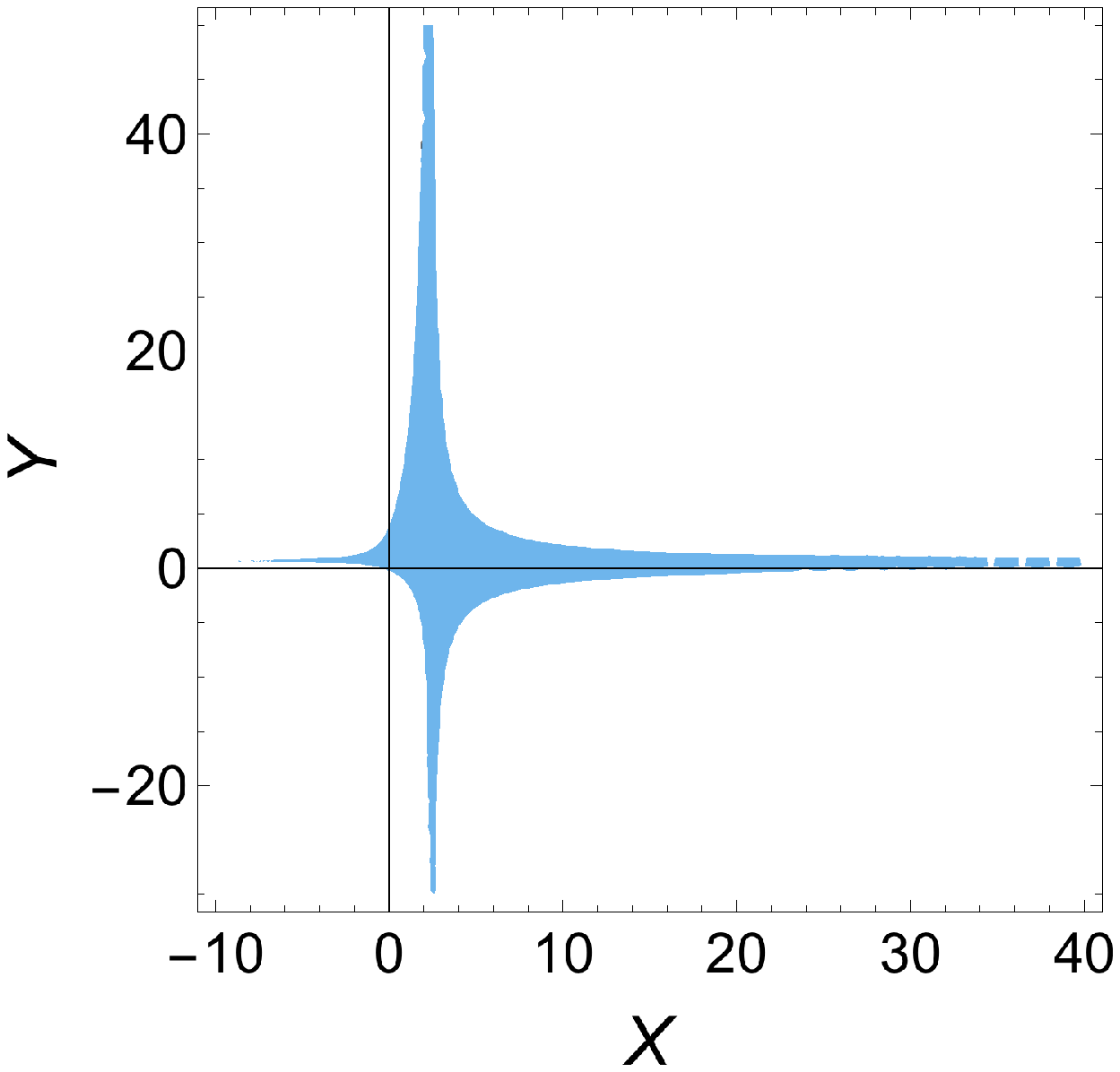}
\caption{The allowed region in the plane $X$ vs $Y$, using the constraint Eq. (\ref{bsg}), which is obtained from the radiative inclusive decay $B\to X_s \gamma$. We obtain the Scenario Ib, which is shown in the left panel, with $0.1\leq \cos (\beta-\alpha) \leq 0.5$, $\chi_{kk}^{u} =1.5$ ($k$=2,3), $\chi_{22}^{d} =1.8$, $\chi_{33}^{d} =1.2$, $\chi_{23}^{u,d} =0.2$, $\chi_{22}^{\ell} =0.5$,   $\chi_{33}^{\ell} =1.2$, $\chi_{23}^{\ell} =0.1$. For Scenario IIa and Y, the allowed region is given in the right panel  with $\cos (\beta-\alpha)=0.1$, $\chi_{22}^{u} =0.5$, $\chi_{33}^{u} =1.4$, $\chi_{22}^{d} =2$, $\chi_{33}^{d} =1.3$, $\chi_{23}^{u} =-0.53$, $\chi_{23}^{d} =0.2$, $\chi_{22}^{\ell} =0.4$,   $\chi_{33}^{\ell} =1.2$, $\chi_{23}^{\ell} =0.1$. For both cases $m_h=125$ GeV, 130 GeV $ \leq m_H\leq 300$ GeV, 100 GeV 
$ \leq m_A\leq 250$ GeV, 110 GeV $ \leq m_{H^\pm}\leq 200$ GeV. }
\label{scan}
\end{figure}

We scan the parameters space of the model and we only consider as viable the points that avoid the 
aforementioned theoretical and experimental bounds and that are fully consistent with the most recent results of Higgs physics from LHC. Taking into account that our model provides interesting new physics in the form of a substantial enhancement of the  decay $\phi \to s\bar{b}+h.c.$  $(\phi = h, \, H)$, as a direct consequence of the off-diagonal terms of the texture of the Yukawa matrices. In our scanning,  we ask that BR$(\phi \to s\bar{b}+h.c.)\sim 0.01$ to $0.1$, keeping the decay $\phi\to b \bar{b}$ dominant. Firstly, we scan the off-diagonal terms of the Yukawa matrices and after we chose some interesting sets of the $\chi$'s parameters, which are consistent with the flavor physics constraints and Higgs physics bounds used in the analysis of \cite{HernandezSanchez:2012eg,Felix-Beltran:2013tra,HernandezSanchez:2013xj}, where we have shown that several meson-physics processes are very sensitive  to  charged Higgs boson exchange, and the off-diagonal terms of Yukawa matrices given in the Eq. (\ref{cheng-sher}) are kept constrained in the following range:
\begin{eqnarray}
-0.06 \leq (\chi_n^d)_{23} \leq 0.3, \,\,\,\,\,\, -0.3 \leq (\chi_n^u)_{23} \leq 0.5.
\end{eqnarray}
 Secondly, we fix the $\chi$'s parameters and the masses of the following Higgs bosons, $m_h=125$ GeV, $m_A =100$ GeV and $m_{H^\pm} =$ 110 GeV. We run the mass $m_H$ of the Higgs boson $H$ from 130 GeV up to 200 GeV. Therefore, we can reduce the study of the parameter space to that of the couplings $X$ and $Y$ only, which are constrained strongly by the inclusive radiative decay $B \to X_s \gamma $ through the following bound:
 \begin{eqnarray}
-1.7 \leq {\rm Re} \Big( \frac{X_{33}Y_{32}}{ V_{tb} V_{Vts}}\Big) \leq 0.3, 
\label{bsg}
\end{eqnarray}
where $X_{33}$, $Y_{32}$ are defined in Eqs. (\ref{Xij})--(\ref{Yij}) and $V_{tb}$ and $V_{ts}$ are elements of the
Cabibbo-Kobayashi-Maskawa (CKM) matrix.  From the constraint in Eq. (\ref{bsg}), we can define the allowed region for two general cases. (a) For the case I defined by: $X=-Y$ or $X=Y$, $0.1\leq \cos (\beta-\alpha) \leq 0.5$, and fixing the parameters of Yukawa matrices, $\chi_{kk}^{u} =1.5$ ($k$=2,3), $\chi_{22}^{d} =1.8$, $\chi_{33}^{d} =1.2$, $\chi_{23}^{u,d} =0.2$, $\chi_{22}^{\ell} =0.5$, $\chi_{33}^{\ell} =1.2$, $\chi_{23}^{\ell} =0.1$. One can see in  Fig. \ref{scan} in the left panel the allowed region for $Y=-X\leq 15$ and for the case $X=Y \leq 20$. This region could represent the case of the 2HDM-I plus a deviation given for the flavor symmetry of the Yuwaka matrices. (b) For the case II given by: $X>>Y$, with $\cos (\beta-\alpha)=0.1$, $\chi_{22}^{u} =0.5$, $\chi_{33}^{u} =1.4$, $\chi_{22}^{d} =2$, $\chi_{33}^{d} =1.3$, $\chi_{23}^{u} =-0.53$, $\chi_{23}^{d} =0.2$, $\chi_{22}^{\ell} =0.4$,   $\chi_{33}^{\ell} =1.2$, we can see that the large values for parameter X is permitted. This case could be the incarnation of both the 2HDM-II and 2HDM-Y (or  flipped model) plus a deviation given by the four-zero texture of the Yukawa matrices. Considering these criteria, we chose three interesting scenarios from the versions of 2HDM-III given in the Tab. \ref{tab:table1}:  {\bf Scenario Ib} which is related to the 2HDM-I-Like, with  $\cos (\beta)= 0.5$,  {\bf Scenario IIa}  is the case 2HDM-II-like, with  $\cos (\beta)= 0.1$,  and {\bf Scenario Ya} is  the case 2HDM-Y-like, with  $\cos (\beta)= 0.1$. 

\subsection{Benchmark scenarios}

 Taking in account the scan of the parameters space,  we chose the benchmark scenarios, where their main features can be recapped as follows:
\begin{itemize}

  \item {\bf Scenario Ib}: 2HDM-III  as 2HDM-I, with the couplings $\phi ff$ given by $g_{\rm 2HDM-III}^{\phi ff}= g_{\rm 2HDM-I}^{\phi ff} + \Delta g$ and $\cos (\beta-\alpha)=0.5$, $\chi_{kk}^{u} =1.5$ ($k$=2,3), $\chi_{22}^{d} =1.8$, $\chi_{33}^{d} =1.2$, $\chi_{23}^{u,d} =0.2$, $\chi_{22}^{\ell} =0.5$,   $\chi_{33}^{\ell} =1.2$, $\chi_{23}^{\ell} =0.1$, $m_A=100$ GeV and $m_{H^\pm}= 110 $ GeV, taking $X$ and $Y$ located in the blue region of the left panel from Figure \ref{scan}.
  \item {\bf Scenario IIa}: 2HDM-III  as 2HDM-II, namely, the couplings $\phi ff$ given by $g_{\rm 2HDM-III}^{\phi ff}= g_{\rm 2HDM-II}^{\phi ff} + \Delta g$ and $\cos (\beta-\alpha)=0.1$, $\chi_{22}^{u} =0.5$, $\chi_{33}^{u} =1.4$, $\chi_{22}^{d} =2$, $\chi_{33}^{d} =1.3$, $\chi_{23}^{u} =-0.53$, $\chi_{23}^{d} =0.2$, $\chi_{22}^{\ell} =0.4$,   $\chi_{33}^{\ell} =1.2$, $\chi_{23}^{\ell} =0.1$, $m_A=100$ GeV and $m_{H^\pm}= 110 $ GeV, taking $X$ and $Y$ allowed in the right panel of the Figure \ref{scan}.
\item {\bf Scenario Y}: 2HDM-III  as 2HDM-Y, namely, the couplings $\phi ff$ given by $g_{\rm 2HDM-III}^{\phi ff}= g_{\rm 2HDM-Y}^{\phi ff} + \Delta g$ and $\cos (\beta-\alpha)=0.1$, $\chi_{22}^{u} =0.5$, $\chi_{33}^{u} =1.4$, $\chi_{22}^{d} =2$, $\chi_{33}^{d} =1.3$, $\chi_{23}^{u} =-0.53$, $\chi_{23}^{d} =0.2$, $\chi_{22}^{\ell} =0.4$,   $\chi_{33}^{\ell} =1.1$, $\chi_{23}^{\ell} =0.1$, $m_A=100$ GeV and $m_{H^\pm}= 110 $ GeV, taking the same $X$ and $Y$ for the Scenario IIa.
\end{itemize}

Hereinafter, we only simulated benchmarks where  $\sigma.BR(\phi \to b\bar s)$ (cross section of the  charged current production $\nu_e \phi q$ multiplied by Branching Ratio of the channel decay $\phi \to b\bar s+$ c.c., with ($\phi = h, \,\, H$)),
are more than 0.15 fb so that, for an integrated luminosity of 
100 fb$^{-1}$, we can start with at least 15 events.  Finally, when producing differential spectra of physical observables,
we will concentrate on three 2HDM scenarios where the number of 
Higgs signal events in the $b\bar s$ + c.c. mode are large enough in order to be able to appreciate  the underlying dynamics.

\section{Numerical Analysis}

In this section, we describe first the production of Higgs signal. We then 
discuss the most important SM backgrounds and the different kinematics selections on the 
simulated events.

\subsection{Higgs bosons signals}

We consider the leading production processes\footnote{The charged-current 
production is approximately 5 times larger than the neutral current 
production. Moreover, the neutral current production contains an electron 
and, since we are vetoing leptons in this particular analysis, we 
consider only the charged processes.} of Higgs boson: $\nu_e \phi q$, 
where $\phi$ = $h$ and $H$ while $q$ is a light-flavor quark (i.e., $u,d,s,c$).
We assume that $\phi$ is dominantly decaying into $b \bar s$  (plus 
charge conjugation). So both of our signals, the lighter Higgs 
as well as the heavier Higgs one, contain three jet (one is forward and 
two are central), missing (transverse) energy and no-lepton. 
Out of the two central jets, one is $b$-tagged and the other is a 
light-flavor jet. We estimated the parton level signal cross sections with flavor-violation 
within the 2HDM-III by using {\tt CalcHEP} \cite{Belyaev:2012qa}. This implementation also calculates the BRs 
of the Higgs boson $\phi$ into $b \bar s$.
For estimating the cross sections at the LHeC \cite{cern:lhec,Bruening:2013bga,AbelleiraFernandez:2012ty,AbelleiraFernandez:2012cc,bm:lhec,lheclumi},
we consider an electron beam, of energy $E_{e^-}$= 60 GeV and a proton beam of energy $E_{p}$= 7000 GeV, 
corresponding to a center-of-mass energy of approximately $\sqrt s = 1.296$ TeV.
The integrated luminosity is 100 fb$^{-1}$.
To estimate the event rates at parton level we applied the following 
basic pre-selections:
\beq \label{presel}
p^q_T >  15~{\rm GeV}, \qquad \Delta R (q,q) > 0.4 \,
\eeq
with $\Delta R = \Delta \eta^2 + \Delta \phi^2$, where $\eta$ and $\phi$ are the 
pseudo-rapidity and azimuthal angle, respectively. 
We take $m_t$=173.3 GeV as the top-quark pole mass. We set the renormalization and 
factorization scale at the $Z$-boson mass (which is approximately the momentum transfer 
scale for the signal) and adopt CTEQ6L \cite{Pumplin:2002vw} as Parton Distribution Functions (PDFs), 
with $\alpha_{\rm s}$ (the strong coupling constant) evaluated 
consistently at all stages (PDFs, hard scattering and decays). 

\begin{figure}[ht!]
\centering
 \includegraphics[width=2.9in]{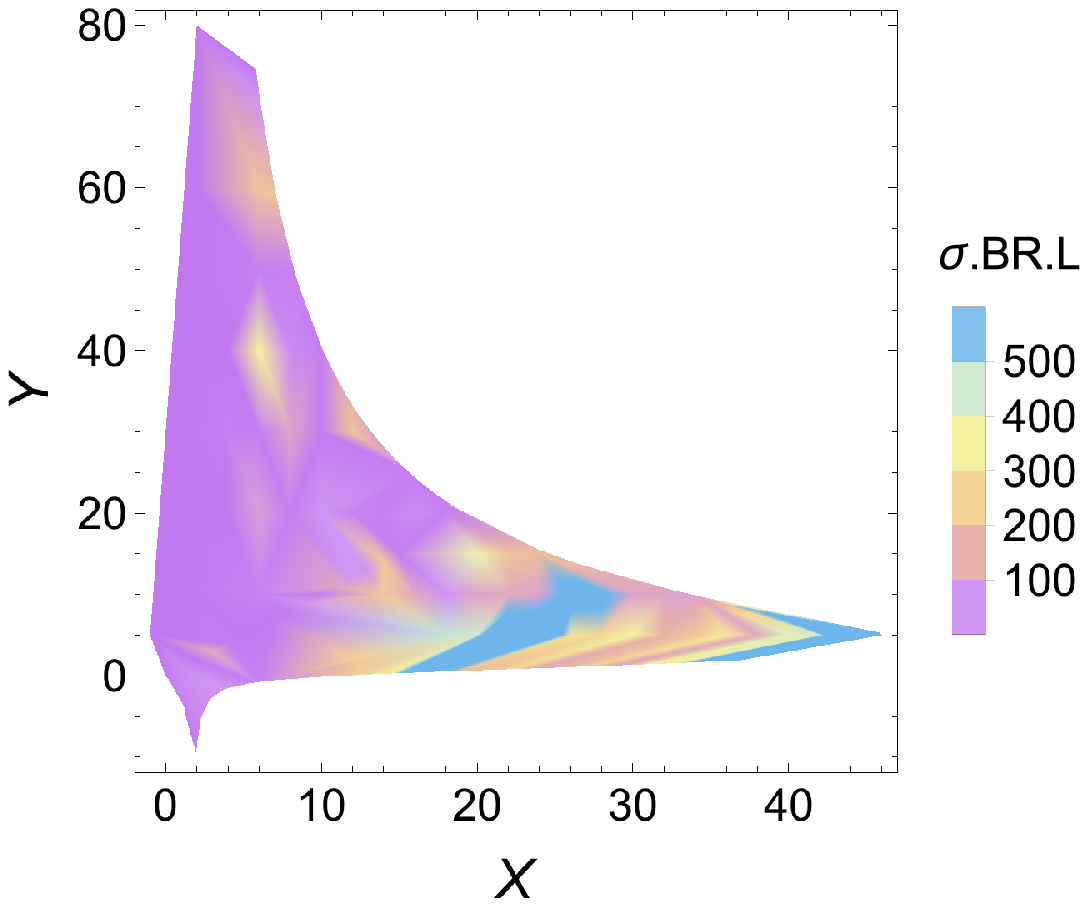}
  \includegraphics[width=2.9in]{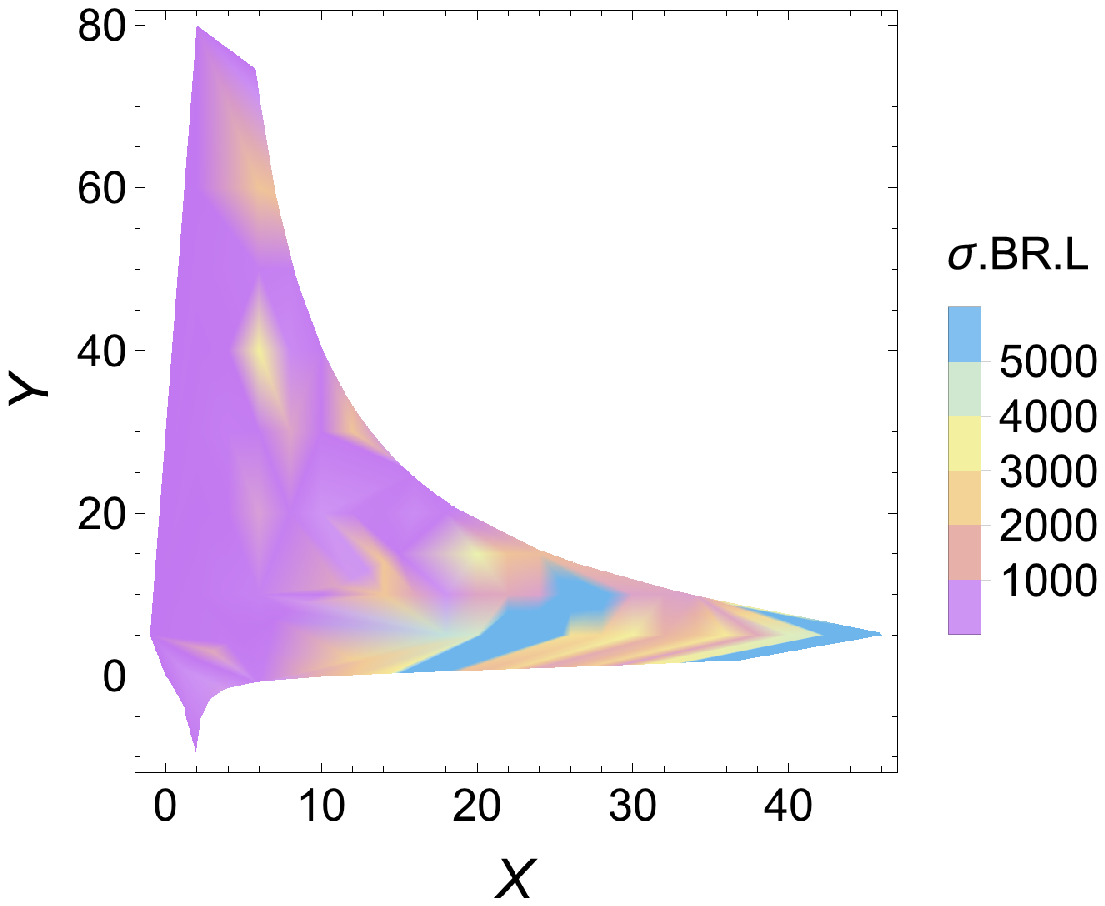}
\caption{Event rates  $(\sigma.BR. L)$ at parton level   for the neutral Higgs boson $h$, where $L$ is the integrated luminosity. We show Scenario Ib for 100 fb$^{-1}$ (left panel) and 1000 fb$^{-1}$ (right panel). We consider $m_h=125$ GeV.}
\label{NEIa}
\end{figure}

\begin{figure}[ht!]
\centering
 \includegraphics[width=2.9in]{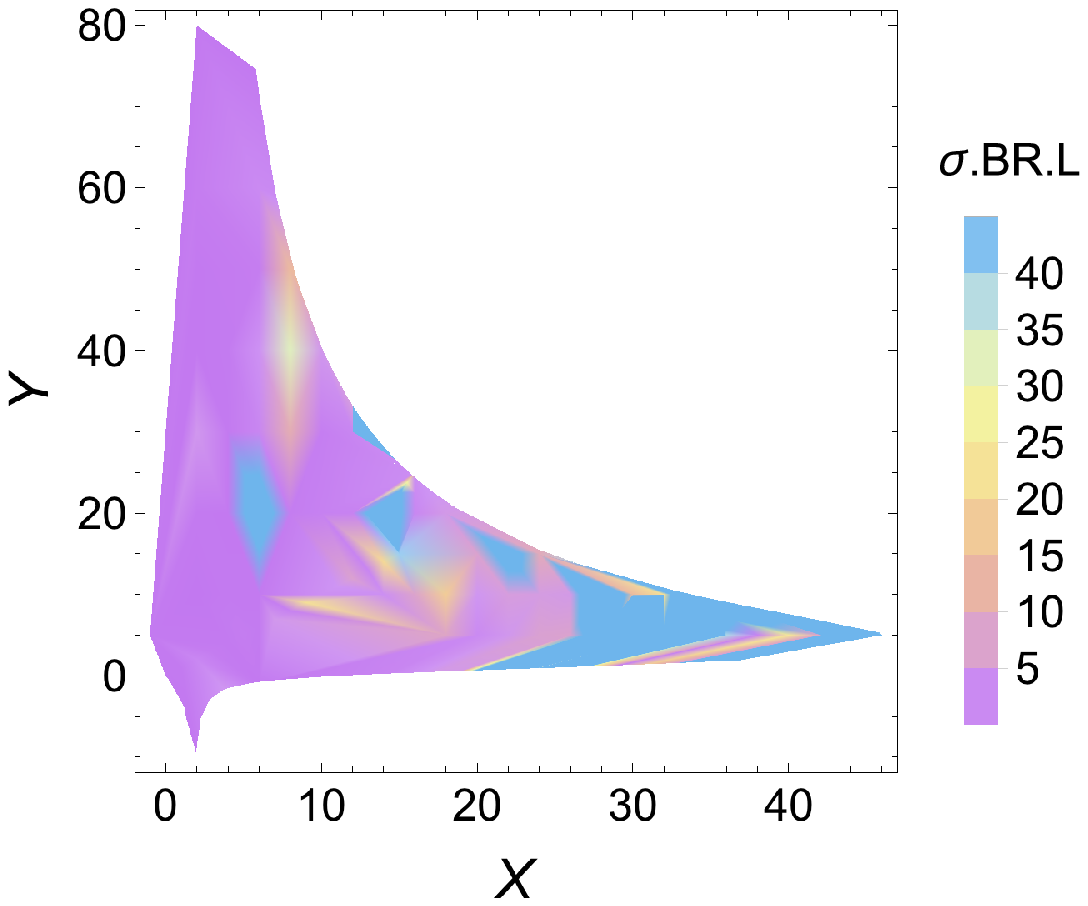}
  \includegraphics[width=2.9in]{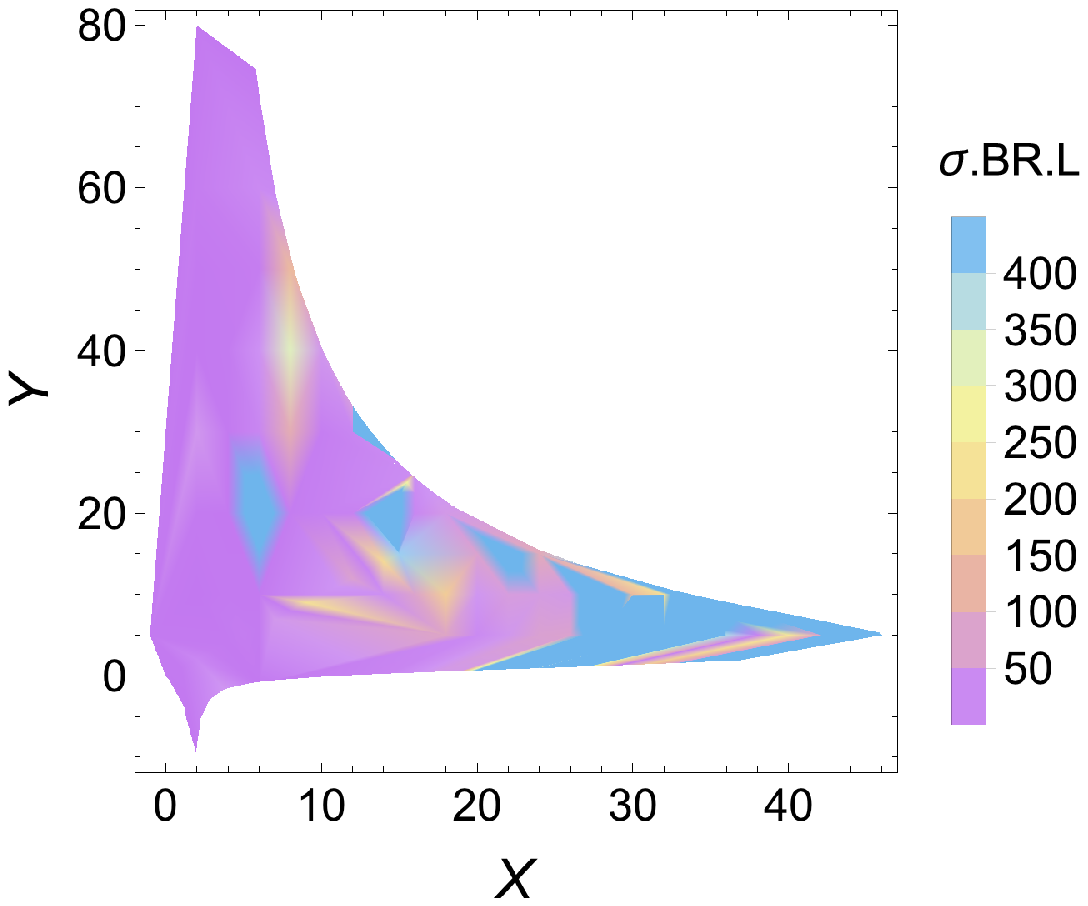}
\caption{Event rates  $(\sigma.BR. L)$ at parton level   for the neutral Higgs boson $H$, where $L$ is the integrated luminosity. We show Scenario Ib for 100 fb$^{-1}$ (left panel) and 1000 fb$^{-1}$ (right panel). We consider $m_H=130$.} 
\label{NEIb}
\end{figure}

\begin{figure}[ht!]
\centering
 \includegraphics[width=2.9in]{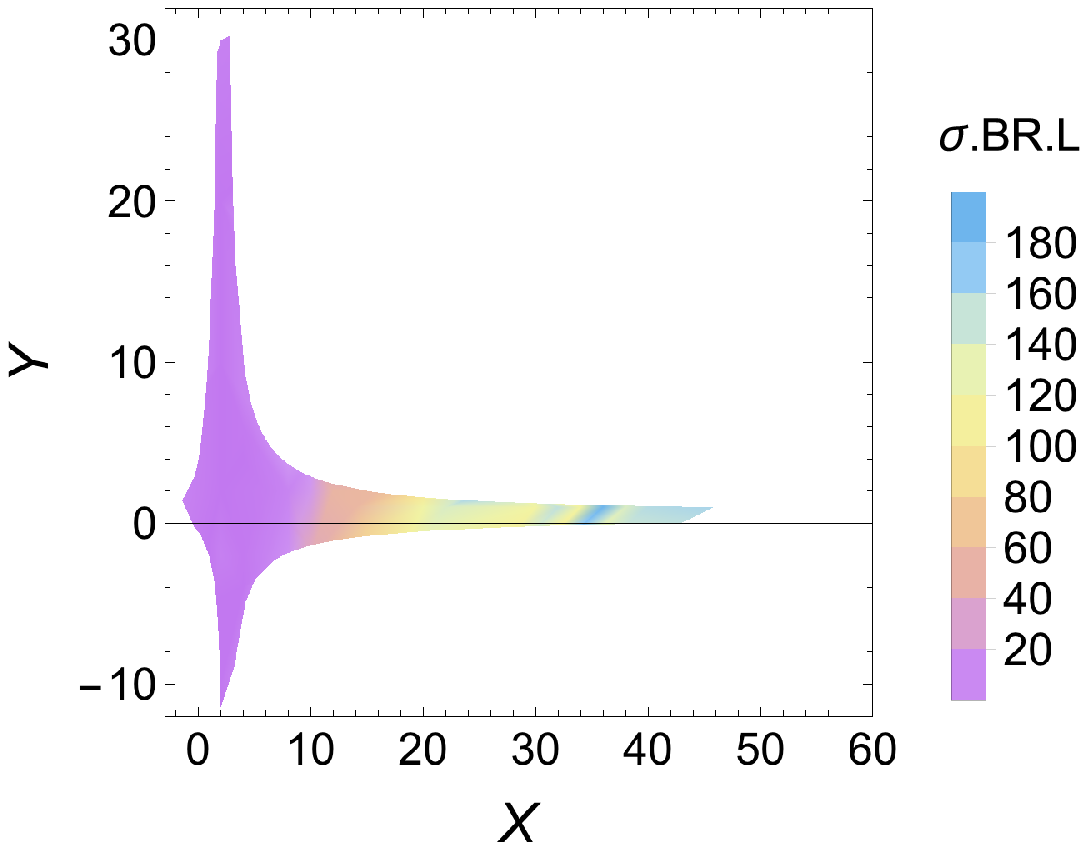}
  \includegraphics[width=2.9in]{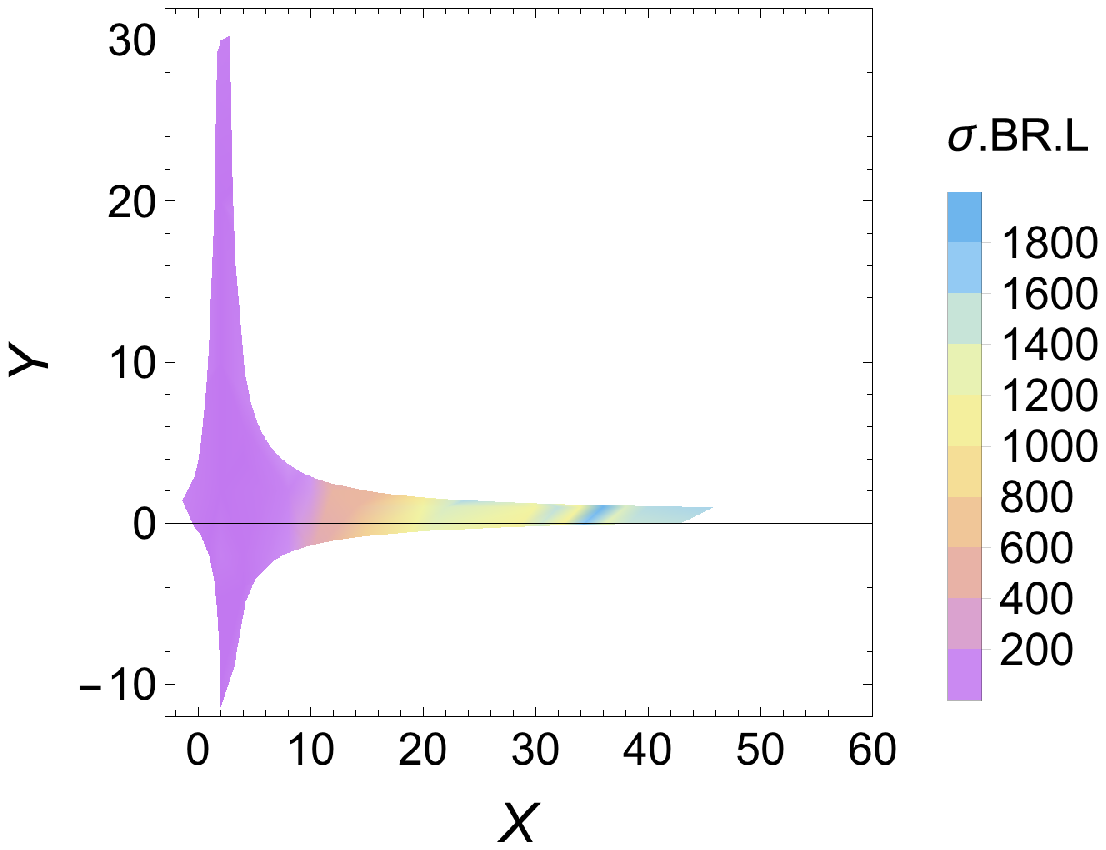}
\caption{Event rates  $(\sigma.BR. L)$ at parton level   for the neutral Higgs boson $h$, where $L$ is the integrated luminosity. We show Scenario IIa for 100 fb$^{-1}$ (left panel) and 1000 fb$^{-1}$ (right panel). We consider $m_h=125$ GeV.}
\label{NEIIa}
\end{figure}

\begin{figure}[ht!]
\centering
 \includegraphics[width=2.9in]{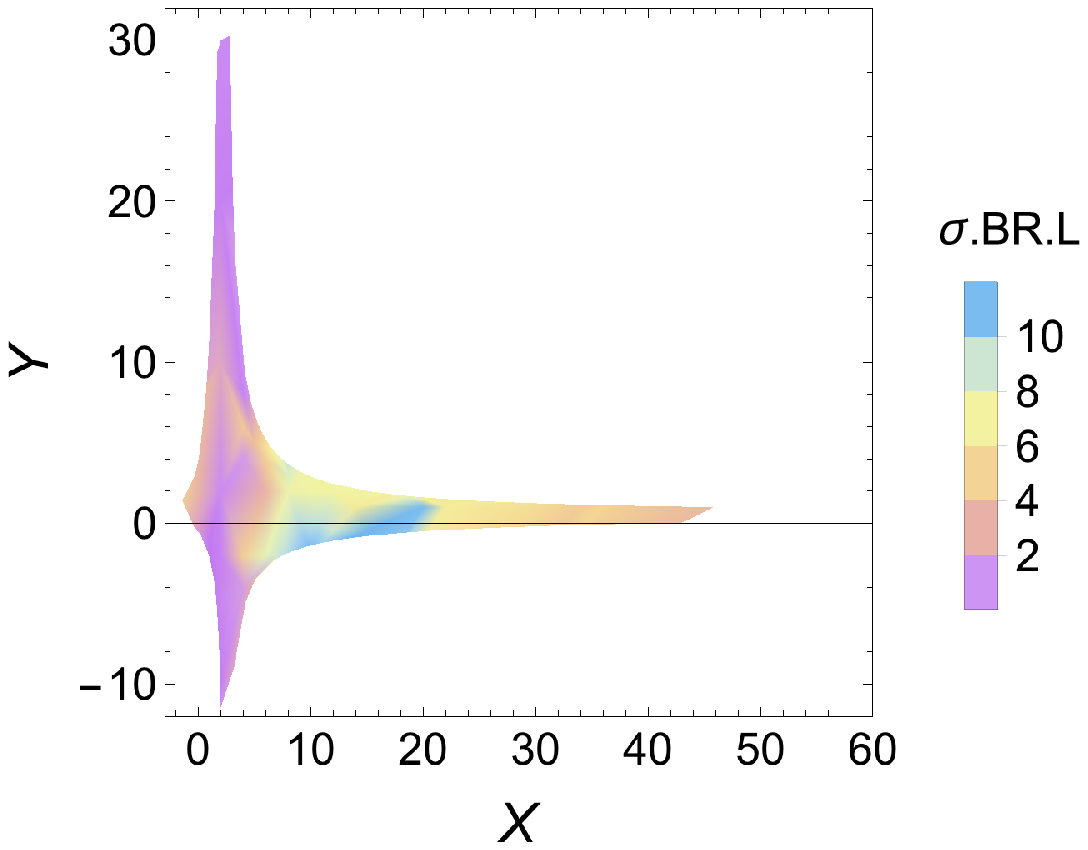}
  \includegraphics[width=2.9in]{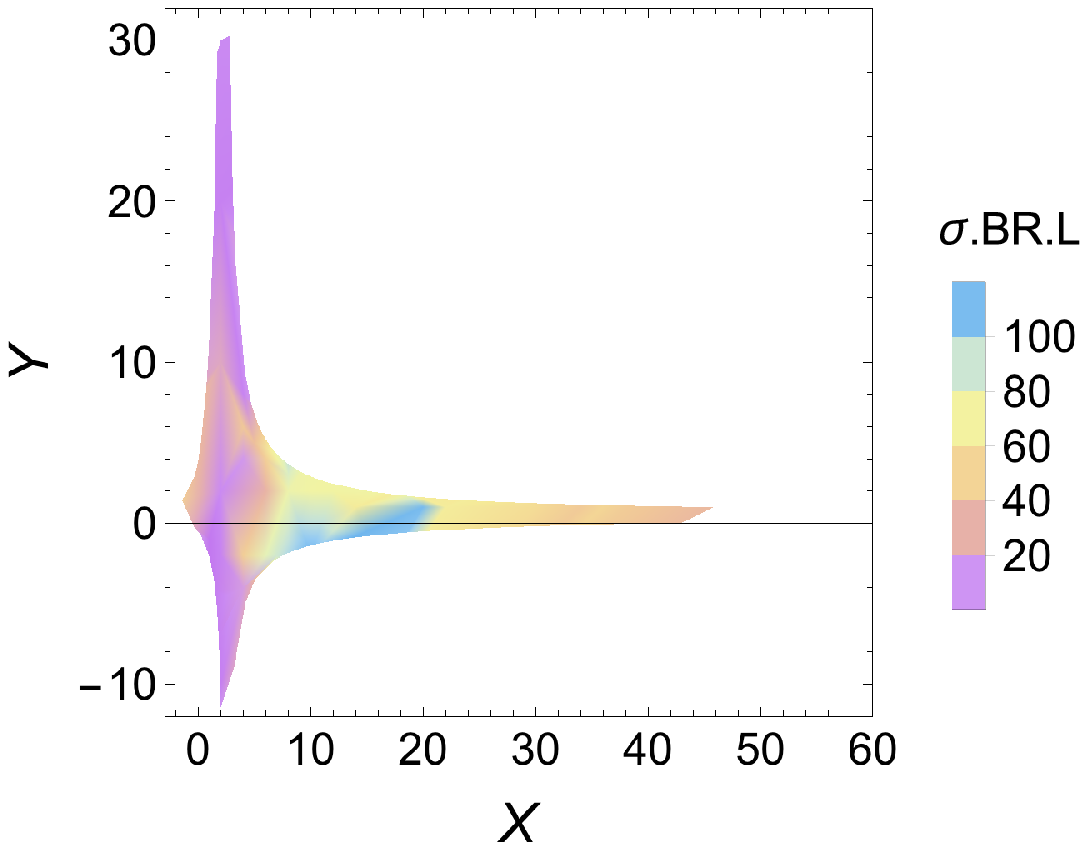}
\caption{Event rates  $(\sigma.BR. L)$ at parton level   for the neutral Higgs boson $H$, where $L$ is the integrated luminosity. We show Scenario IIa for 100 fb$^{-1}$ (left panel) and 1000 fb$^{-1}$ (right panel). We consider $m_H=130$.} 
\label{NEIIaH}
\end{figure}

\begin{figure}[ht!]
\centering
 \includegraphics[width=2.9in]{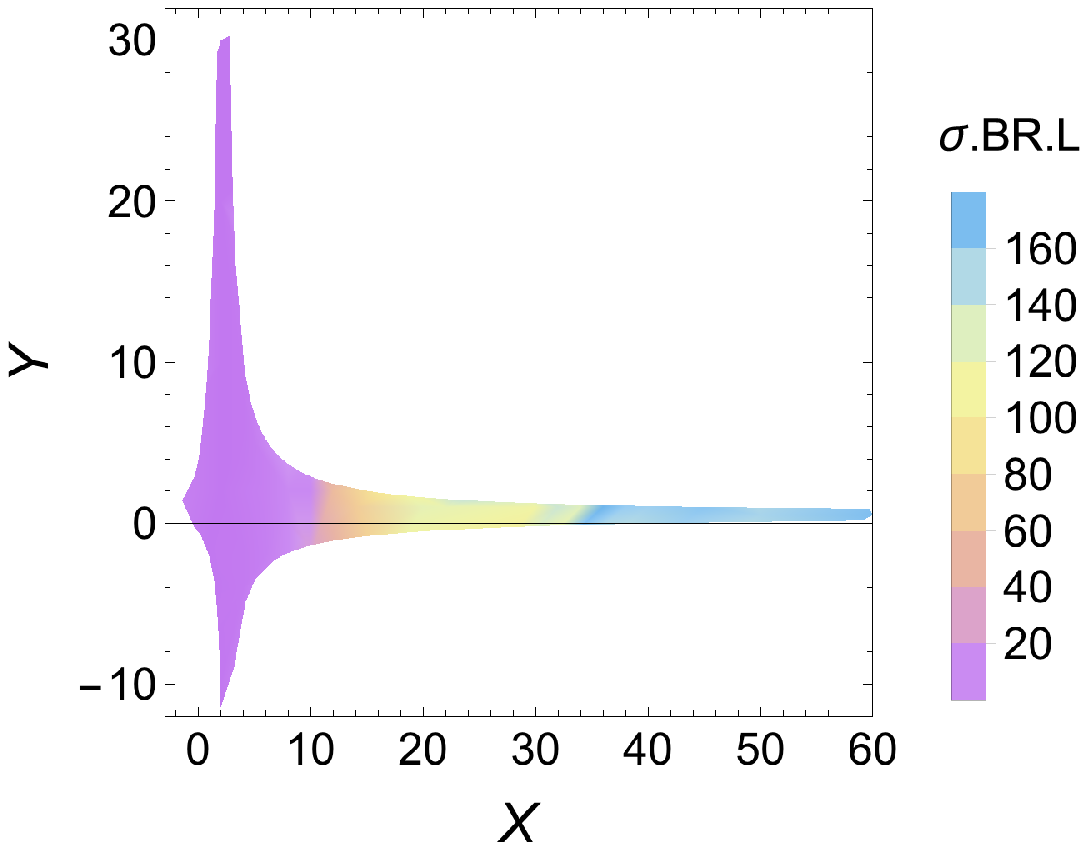}
  \includegraphics[width=2.9in]{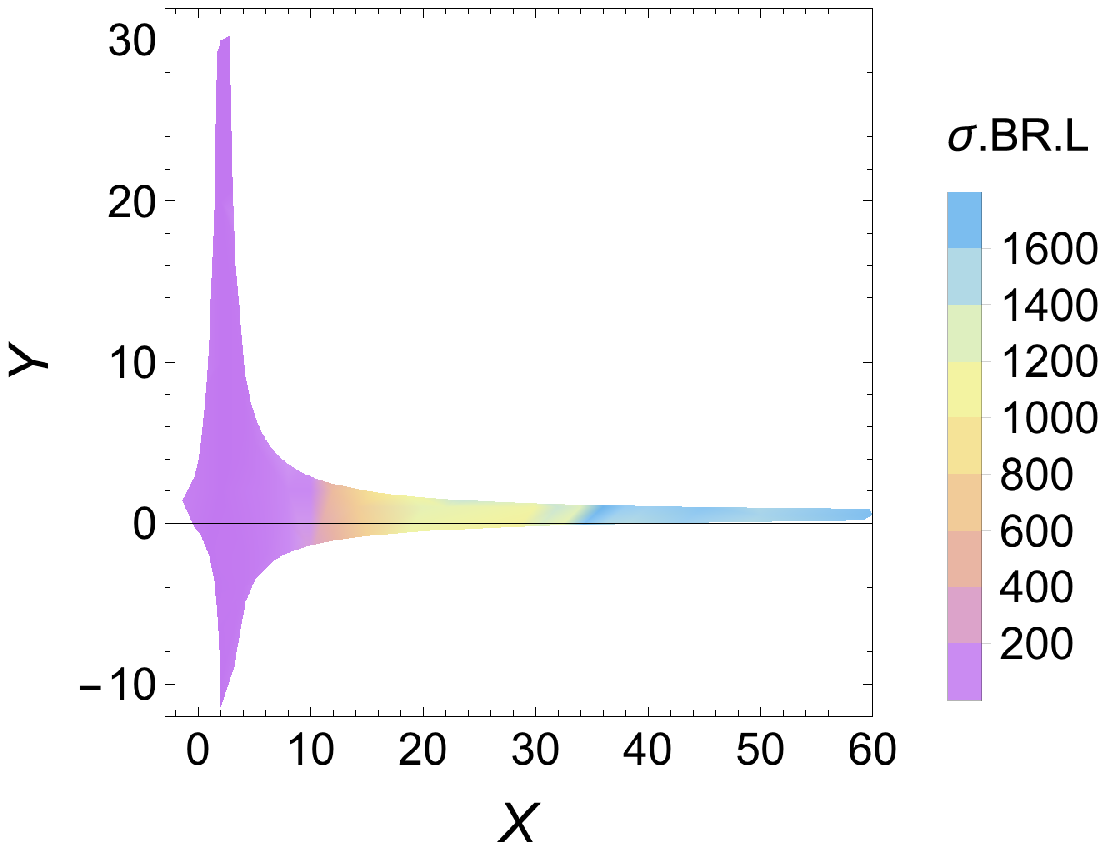}
\caption{Event rates  $(\sigma.BR. L)$ at parton level   for the neutral Higgs boson $h$, where $L$ is the integrated luminosity. We show Scenario Y for 100 fb$^{-1}$ (left panel) and 1000 fb$^{-1}$ (right panel). We consider $m_h=125$ GeV.}
\label{NEYa}
\end{figure}

\begin{figure}[ht!]
\centering
 \includegraphics[width=2.9in]{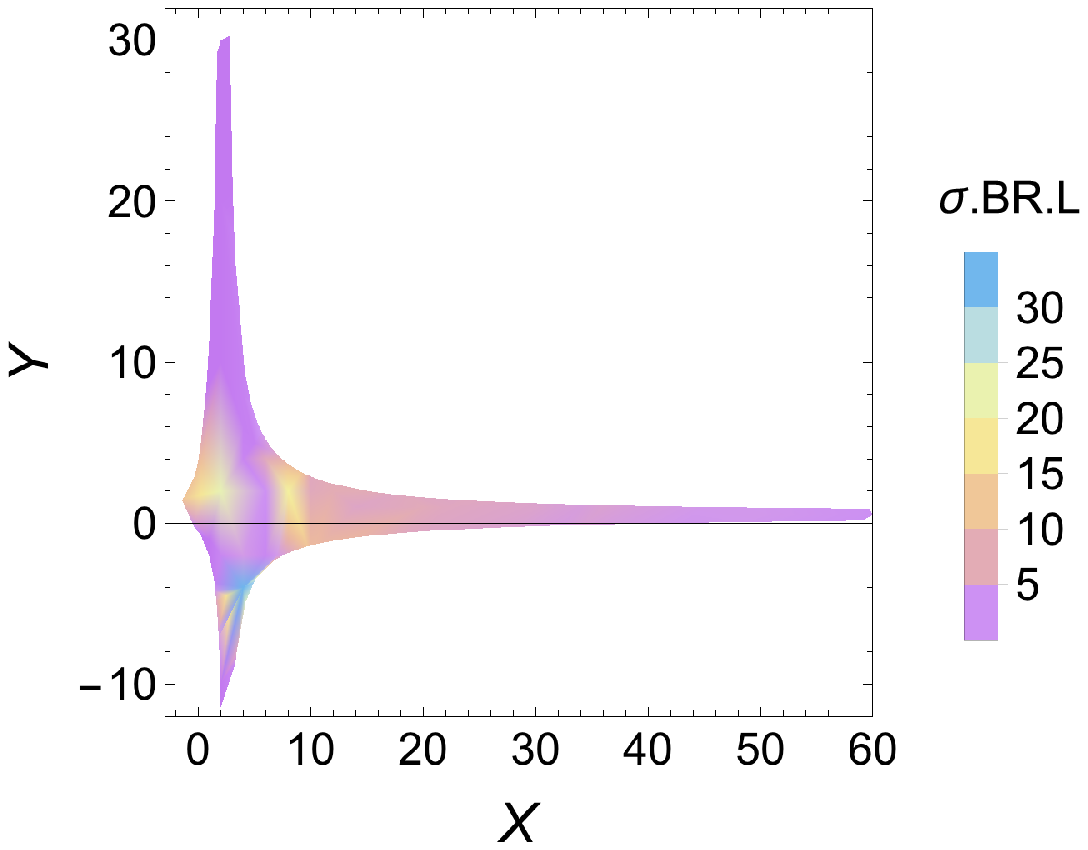}
  \includegraphics[width=2.9in]{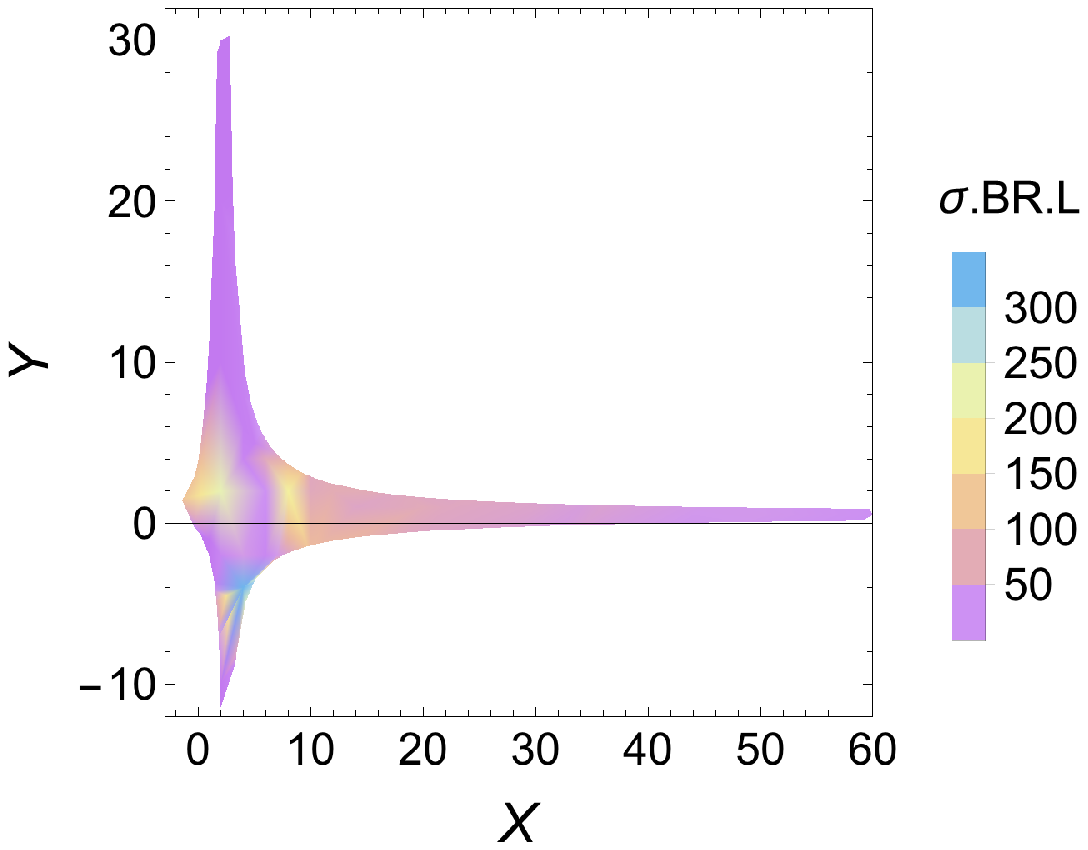}
\caption{Event rates  $(\sigma.BR. L)$ at parton level   for the neutral Higgs boson $H$, where $L$ is the integrated luminosity. We show Scenario Y for 100 fb$^{-1}$ (left panel) and 1000 fb$^{-1}$ (right panel). We consider $m_H=130$.} 
\label{NEYaH}
\end{figure}

  Considering the latter, we calculate in the allowed regions given above in the Figure \ref{scan}, the event rates  $(\sigma.BR. L)$ at parton level   for the neutral Higgs bosons $h$ and $H$ in the Scenarios Ib, IIa and Y, respectively, considering both  luminosities of 100 fb$^{-1}$ (left panel) and 1000 fb$^{-1}$ (right panel), which are shown in the Figures \ref{NEIa}--\ref{NEYaH}. One can see that the blue region contains the best benchmark points for all scenarios. We show that the most optimistic is in fact Scenario Ib for both Higgs bosons $h$ and $H$, which reach events rates of order 500--1300 (5000-13000) with an integrated luminosity of 100 fb$^{-1}$ (1000 fb$^{-1}$), although Scenarios IIa and Y also have some interesting benchmark points where one can obtain 17 events rates at the same luminosity.   The Tab. \ref{tab:sigmabr} shows the benchmark points that we select as interesting for studies at the LHeC. There are twenty-seven in total, obtained by taking the same three different values of the $H$ mass ($m_H=$ 130, 150, 170 GeV) in correspondence to nine different configurations of the other parameters.The product of cross sections times the relevant BRs ($\sigma.bs$) are shown in Tab. \ref{tab:sigmabr}. 
\begin{table}[!t]
\centering
{\scriptsize
\begin{tabular}{|c|c|c|c|c|c|c|c|c|c|c|c|}
\hline
\hline
2HDM&$X$&$Y$&$Z$&\multicolumn{2}{c|}{$m_h=125$ GeV}&\multicolumn{2}{c|}{$m_H=130$ GeV} &\multicolumn{2}{c|}{$m_H=150$ GeV} &\multicolumn{2}{c|}{$m_H=170$ GeV}\\
&&&&bs&$\sigma.bs$ &bs&$\sigma.bs$ &bs&$\sigma.bs$ &bs&$\sigma.bs$\\ 
\hline
\hline
Ib35&28&10&28&15.66&6.392&51.8&1.209& 51.6&0.30&1.58&0.117\\
Ib47&$30$& $5$&$30$&16.14&3.086&48.2&10.983& 48.0&0.127&1.80&0.839\\
Ib57&44&5&44&17.58&11.861&38.6&5.14& 38.4&2.303&3.68&0.137\\
\hline
IIa11&20&2&20&1.42&1.055&25.2&0.097& 25.0&0.091&24.8&0.085\\
IIa14&$26$&$2$&$26$&1.44&1.651&26.0&0.059& 25.8&0.054&25.6&0.049\\
IIa26&36&1&36&1.46&1.621&26.4&0.045&
26.2&0.042&26.0&0.038\\
\hline
Ya11&20&2&$-2$&1.42&1.084&25.2&0.062& 25.0&0.059&24.8&0.054\\
Ya12&$22$&$2$&$-2$&1.44&1.078&25.6&0.057& 25.4&0.053&25.2&0.048\\
Ya14&26&2&$-2$&1.46&1.441&26.0&0.057& 25.8&0.053&25.6&0.049\\
\hline
\hline
\end{tabular}
\centering
\caption{Parameters for few optimistic benchmark points in the 2HDM-III as a 2HDM-I, -II and -Y configuration.
Here {\it bs} stands for BR($\phi \to b \bar s + \bar b s$), in units of $10^{-2}$,
where $\phi= h, H$,  while $\sigma$.$bs$ stands for the 
cross section multiplied by the above BR as obtained at the LHeC  in units
of fb. We have analyzed only the benchmarks where the $\sigma.bs$ is greater 
than 0.15 fb, so that at least 15 events are produced  for 100 fb$^{-1}$.
} \label{tab:sigmabr}
}
\end{table}

Restricting ourselves to the points for which the inclusive event rates are most optimistic, all estimated by taking all the light-flavor quarks,  the $b$-quark and the gluon as fluxes inside the proton and upon considering  appropriate flavor-mixing where appropriate,
we have then proceeded as follows. The top-quark and $W$-boson were allowed to 
decay freely as implemented in {\tt PYTHIA} \cite{pythia}. Following this, it was recognized that the signal processes have unique kinematic profiles. In particular, the final state quark transverse momentum is less than the mass of the vector bosons,
its energy is very high with a small angle with respect to the beam directions (i.e., high forward rapidity).
This will serve as guidance in our event selection. However, before proceeding further in this direction, we have to acknowledge at this point that 
these processes and their kinematic features to discover generic Higgs 
bosons have been studied since long \cite{epold,epold2}. Further 
parton level studies have been performed within the SM recently in \cite{Han:2009pe}. 
In the context of BSM physics,  cross section estimates while taking into account
Next-to-Leading Order (NLO) correction 
factors have been performed in \cite{Zhe:2011yr} but no signal and background studies have been reported therein.
In fact, dedicated simulations at the event generator level have not been done extensively
and we focus here on these aspects, most importantly, with the intent of detecting  two Higgs bosons
simultaneously in novel flavor-violating modes.

\subsection{Backgrounds}

There are mainly two groups of SM noise to our Higgs signals. 
The charged-current backgrounds, $\nu t \bar b$, $\nu b \bar b j$, $\nu b2j$, $\nu 3j$, 
and the photo-production ones, $e^{-} b \bar b j$, $e^{-} t \bar t$.
For estimating the cross sections of these SM backgrounds, we have used the same 
pre-selections like for the signal, Eq. (\ref{presel}), and identical conventions 
and parameter sets. The expected number of events for 100 fb$^{-1}$ of integrated 
luminosity are given in the third column of Tab. \ref{tab:h125}. 

\subsection{Signal-to-background analysis}

We passed the {\tt CalcHEP} v3.4.7 \cite{Belyaev:2012qa} generated parton level 
event on to {\tt PYTHIA} v.6.408 \cite{pythia}, which handles the parton shower 
(both initial and final), hadronization, heavy hadron decays etc.  The final 
state radiation smears the four-momentum of the jets, thus the invariant 
mass of the Higgs boson signal is less than the actual 
values considered in the event. We also took the experimental resolutions of 
the jet angles and energy using the toy calorimeter {\tt PYCELL}, in accordance 
with the LHeC detector parameters, given in {\tt PYTHIA}.
This has some non-trivial effect since we used the invariant mass to isolate 
the Higgs signal.
In our study we considered the LHC type calorimeter for the LHeC. Although 
in reality this is not the case, for example, unlike ATLAS and CMS, 
the electro-magnetic and the hadronic calorimeter at the LHeC is not 
symmetric. However, since we are not doing detector simulation and 
also we are not considering cracks in the detectors, we applied symmetric 
large rapidity coverage for jets and leptons in our analysis. 
We expect that these assumptions hardly alter our findings.
The detector parameters in the toy calorimeter module {\tt PYCELL} are set according 
to the LHeC detector \cite{AbelleiraFernandez:2012cc}. Specifically, we assume large 
calorimeter coverage $|\eta| < 5.5$, with segmentation (the number of division in $\eta$ and
 $\phi$ are 320 and 200 respectively) $\Delta \eta \times \Delta \phi = 0.0359 \times 0.0314$. 
Further, we have used Gaussian energy resolution \cite{Bruening:2013bga} for both leptons 
($\ell=e,\mu$) and jets (labelled as $j$), with
\beq \label{lhc_res}
{{\Delta E \over E} = { a \over \sqrt{E} } \oplus b},
\eeq
where $a=0.32$, $b=0.086$ for jets and $a=0.085$, $b= 0.003$ for leptons and 
$\oplus$ means addition in quadrature. 
We have used a cone algorithm for the jet-finding algorithm, 
with jet radius $\Delta R(j) = \sqrt{\Delta\eta^{2}+\Delta\phi^{2}} = 0.5$. 
Calorimeter cells with $E_{T,min}^{\rm cell} \ge 5.0$ GeV are considered
to be potential candidates for jet initiators. All cells with $
E_{T,\rm min}^{\rm cell} \ge 1.0$ GeV were treated as part of the would--be jet. 
A jet is required to have minimum summed $E_{T,min}^{j} \ge 15$ GeV
and the jets are ordered in $E_{T}$.
Leptons ($\ell = e, ~\mu$) are selected if they satisfy the requirements:
$E_T^{\ell} \ge 15$ GeV and $\left| \eta^{\ell} \right| \le 3.0$. 
In our jet finding algorithm we include leptons as parts of jets. 
Finally we separate them, putting some isolation criterion as follows: 
if we find a jet near a lepton, with $\Delta R (j-\ell) \le 0.5$ and $ 0.8 \le
E_{T}^{j}/E_{T}^{\ell} \le 1.2$, i.e. if the jet $E_T$ is nearly
identical to that of this lepton, the jet is removed from the list of
jets and treated as a lepton. However, if we find a jet within $\Delta
R (j-\ell) \le 0.5$ of a lepton, whose $E_T$ differs significantly
from that of the lepton, the lepton is removed from the list of leptons.
This isolation criterion mostly remove leptons from $b$ or $c$ decays.
We reconstructed the missing (transverse) energy ($\met$) from all observed particles
and shown in left panel of Fig. \ref{metetaf}. We have also calculated the 
same from the energy deposition in the calorimeter cells and found consistency 
between these two methods. 
Only jets with $|\eta^j| < 2.5$ and  $E_{T}^j \geq 15$ GeV
``matched'' with a $b-$flavored hadron ($B-$hadron), i.e.  with
$\Delta R(j,B-{\rm hadron}) < 0.2$ is considered to be ``taggable''. 
We assume that these jets are actually tagged with probability 
$\eps_b = 0.50$. 
We also adopted mis-tagging of non$-b$ jets as $b-$jets and treated $c-$jets 
differently from the gluon and light-flavor jets. A jet with 
$\left| \eta^j \right| \le 2.5$ and $E_{T}^j \geq 15$ GeV matched 
with a $c-$flavored hadron ($C-$hadron, e.g., a $D-$meson or $\Lambda_c-$baryon), 
i.e., with $\Delta R(j,C-{\rm hadron}) < 0.2$, is again
considered to be taggable, with (mis-)tagging probability $\eps_c = 0.10$.
Jets that are associated with a $\tau-$lepton, with $\Delta R(j,\tau) \le 0.2$, and all jets
with $\left| \eta^j \right| > 2.5$, are taken to have vanishing tagging
probability. All other jets with $E_{T}^j \geq 15$ GeV and
$\left| \eta^j \right| \leq 2.5$ are assumed to be (mis-)tagged with
probability $\eps_{u,d,s,g} = 0.01$. These efficiencies follow 
recent LHeC analyses~\cite{Han:2009pe}.

\begin{table}[t!]
\centering
{\scriptsize
\begin{tabular}{|c||c|c||c|c|c|c|c|c|c|c|c||c|}
\hline
&&&&&&&&&&&&\\
Proc&SimEvt&RawEvt& a & b & c & d& e &f&g&h&i&${\cal S}$\\ 
\hline
\hline
Ib35&100K&639.2  &447.6 &177.3 &117.1 & 97.4 & 93.8 &37.8 &31.7 &25.4 &15.8 &1.2(3.8)\\
Ib47&100K&308.6  &216.8 & 85.1 & 56.2 & 47.1 & 45.5 &18.4 &15.6 &13.0 & 8.1 &0.62(2.0)\\
Ib57&100K&1186.1 &833.7 &325.7 &215.5 &180.6 &173.9 &70.3 &59.1 &49.3 &31.1 &2.4(7.5)\\
\hline
IIa11&100K& 105.5& 74.3&    29.1&        19.2& 16.0& 15.4& 6.3& 5.3&4.4& 2.8 &0.21(0.70)\\
IIa14&100K& 165.1& 116.1&   45.2&        30.0& 25.4& 24.4& 9.7& 8.3&6.9& 4.4 &0.33(1.05)\\
IIa26&100K& 162.1& 114.4&   44.7&        29.5& 24.5& 23.6& 9.5& 8.1&6.8& 4.3 &0.33(1.03)\\
\hline
Ya11&100K& 108.4&        76.3&        29.8&   19.6& 16.4& 15.8&  6.4& 5.4& 4.6& 2.9&0.22(0.70)\\
Ya12&100K& 107.8&        76.2&        29.6&   19.5& 16.3& 15.7&  6.3& 5.4& 4.5& 2.8&0.21(0.67)\\
Ya14&100K& 144.1&       101.7&        39.8&   26.0& 21.7& 20.8&  8.2& 7.0& 5.9& 3.8&0.29(0.92)\\
\hline
\hline
$\nu t\bar b$&100K&50712.1&28338.4&15293.7&9845.0& 8144.2&7532.7& 2982.1& 2058.0&652.2&139.6&\\
$\nu b\bar bj$&560K&14104.6&6122.8&3656.7& 1858.5&  1787.1&   1650.1& 257.5& 152.5&85.2&15.1&\\
$\nu b2j$&90K&18043.1&8389.2&3013.0& 1691.5& 1445.5&   1373.7& 389.5& 206.1&77.2&11.3&$B$=170.8\\
$\nu 3j$&300K& 948064.2&410393.4& 15560.9& 0.0& 0.0& 0.0& 0.0& 0.0& 0.0& 0.0&$\sqrt{B}$=13.1\\
$e b\bar b j$&115K&256730.1&55099.8& 36353.6&12659.8& 1432.0&200.7& 54.1&24.8&18.0&4.5&\\
$e t\bar t$ &130K& 783.3& 685.0& 384.5& 265.9& 179.3& 26.2& 11.6& 10.5& 3.9& 0.3&\\
\hline
\end{tabular}
}
\caption{Expected number of events after different combinations of
  cuts for signal and backgrounds at the LHeC with 100 fb$^{-1}$
  integrated luminosity for $m_h$=125 GeV. SimEvt stands for the 
  actual number of events analyzed in the Monte Carlo simulations. 
  RawEvt stands for the number of events with only the generator--level cuts 
  (\ref{presel}) imposed; for the signal as well as for background, these are 
  calculated from the total cross section times branching ratio. In the final 
  column we mention the significances(${\cal S}$) defined as ${\cal S} = S / \sqrt{B}$, 
  where signal events $S$, background events $B$ for 100 fb$^{-1}$ of data after all cuts 
  mentioned in the ``i'' column. The number in the parenthesis in the final column 
  represent the significances for 1000 fb$^{-1}$.
 }
\label{tab:h125}
\end{table}

\begin{table}[t!]
\centering
{\scriptsize
\begin{tabular}{|c||c|c||c|c|c|c|c|c|c|c|c||c|}
\hline
&&&&&&&&&&&&\\
Proc&SimEvt&RawEvt& A & B & C & D& E &F&G&H&I&${\cal S}$\\ 
\hline
\hline
Ib35&100K&  120.9& 87.1& 34.1& 26.9& 22.5& 21.6& 7.5&  6.1&   5.3& 3.4&0.28(0.88)\\
Ib47&100K& 1098.3& 790.3& 307.1&  243.9&  204.6& 195.7& 68.5& 56.1& 48.6& 31.3&2.6(8.1)\\
Ib57&100K&  514.0& 371.2&       144.8&       115.0& 96.0&92.0&   31.7& 25.8&22.7& 14.3&1.2(3.7)\\
\hline
IIa11&100K&  9.7&   6.8&   2.7&   2.1& 1.8& 1.7&  0.6&  0.4& 0.3& 0.2&0.02(0.05)\\
IIa14&100K&  5.9&   4.2&   1.7&  1.3&  1.1& 1.0&  0.4&  0.3& 0.2& 0.1&0.01(0.02)\\
IIa26&100K& 4.5&  3.1&     1.3&    1.0&   0.8& 0.8& 0.3& 0.2& 0.1&0.1&0.01(0.02)\\
\hline
Ya11&100K&   6.2& 4.4& 1.8& 1.4& 1.1&1.1& 0.4&  0.3&  0.2& 0.1&0.01(0.02)\\
Ya12&100K&   5.7& 4.0& 1.6& 1.3&1.0& 1.0&0.3& 0.2& 0.2&  0.1&0.01(0.02)\\
Ya14&100K&   5.7& 4.0& 1.6& 1.3&1.0& 1.0& 0.3&0.2& 0.2& 0.1&0.01(0.02)\\
\hline
\hline
$\nu t \bar b$& 100K&50712.1& 28338.4& 15293.7& 10976.4& 9092.4& 8393.6& 2550.9& 1565.5&617.9& 113.7&\\
$\nu b\bar bj$&560K& 14104.6&6122.8& 3656.7& 2145.5& 2062.1& 1902.9& 266.6& 141.0& 87.5& 14.4&\\
$\nu b2j$&90K& 18043.1& 8389.2& 3013.0& 2053.6& 1734.0& 1650.1& 402.8& 143.7& 64.5& 8.1&$B$=147.8\\
$\nu 3j$&300K& 948064.2&410393.4& 15560.9& 0.0& 0.0& 0.0& 0.0& 0.0& 0.0& 0.0&$\sqrt{B}$=12.2\\
$e b\bar b j$&115K& 256730.1&55099.8&36353.6&16838.4& 1826.6& 284.1& 56.4& 31.6& 22.6& 11.3&\\
$e t\bar t$&130K& 783.3&  685.0&384.5& 280.8& 190.8& 27.8& 10.9& 9.3& 3.9& 0.3&\\
\hline
\end{tabular}
}
\caption{Same as Tab. \ref{tab:h125} but for $m_H$=130 GeV. The criterion 
for jets and $b$-tagging are the same, so that the number of events 
in column $A$ and $B$ are the same for all SM backgrounds.}
\label{tab:H130}
\end{table}

\begin{table}[t!]
\centering
{\scriptsize
\begin{tabular}{|c||c|c||c|c|c|c|c|c|c|c|c||c|}
\hline
&&&&&&&&&&&&\\
Proc&SimEvt&RawEvt& A & B & C & D& E &F&G&H&I&${\cal S}$\\ 
\hline
\hline
Ib35&100K& 30.0& 23.3&   9.1&  8.2&  6.9& 6.5&1.5&1.3& 1.2& 0.8&0.10(0.33)\\
Ib47&100K& 12.7&  9.9&   3.8&  3.4& 2.9&2.7&0.6&0.5& 0.5& 0.3&0.04(0.12)\\
Ib57&100K& 230.3& 179.6& 69.3& 62.6& 52.6&49.9&11.7&10.1&9.1&6.4&0.83(2.62)\\
\hline
IIa11&100K& 9.1& 6.9&  2.7&         2.4& 2.0&1.9&0.4&  0.4&0.3&0.2&0.026(0.08)\\
IIa14&100K& 5.4& 4.1& 1.6&  1.4&1.2& 1.1&0.3& 0.2& 0.2& 0.1&0.013(0.04)\\
IIa26&100K&         4.2&  3.2& 1.3& 1.1& 0.9& 0.9&0.2& 0.1&0.1& 0.1&0.013(0.04)\\
\hline
Ya11&100K& 5.9& 4.5&     1.8&         1.6&         1.3& 1.2&  0.3&0.2&0.2& 0.1&0.013(0.04)\\
Ya12&100K&         5.3&         4.0&         1.6&1.4&1.2& 1.1& 0.3&  0.2&0.2& 0.1&0.013(0.04)\\
Ya14&100K&  5.3&  4.0& 1.6&  1.4&1.2& 1.1&0.3&         0.2&0.2&0.1&0.013(0.04)\\
\hline
\hline
$\nu t \bar b$& 100K& 50712.1& 28338.4& 15293.7& 11810.9& 9808.7& 9039.0& 751.7& 476.8& 194.5& 32.3&\\
$\nu b\bar bj$&560K&14104.6& 6122.8& 3656.7& 2395.6& 2300.1&2120.8&  199.3& 112.4&  70.8& 12.4&\\
$\nu b2j$&90K& 18043.1& 8389.2& 3013.0&2427.2& 2030.3& 1933.1& 234.2&  83.7& 41.0& 6.3&$B$=60.1\\
$\nu 3j$&300K& 948064.2&410393.4& 15560.9& 0.0& 0.0& 0.0& 0.0& 0.0& 0.0& 0.0&$\sqrt{B}$=7.7\\
$e b \bar b j$&115K&256730.1& 55099.8& 36353.6& 21280.9& 2270.8& 385.6& 36.1& 24.8& 20.3& 9.0&\\
$e t\bar t$&130K& 783.3& 685.0& 384.5& 291.5& 199.0& 29.1&  3.5& 3.0& 1.2& 0.1&\\
\hline
\end{tabular}
}
\caption{Same as Tab. \ref{tab:H130} but for $m_H$=150 GeV.}
\label{tab:H150}
\end{table}

\begin{table}[t!]
\centering
{\scriptsize
\begin{tabular}{|c||c|c||c|c|c|c|c|c|c|c|c||c|}
\hline
&&&&&&&&&&&&\\
Proc&SimEvt&RawEvt& A & B & C & D& E &F&G&H&I&${\cal S}$\\ 
\hline
\hline
Ib35&100K& 11.7&  9.6&   3.7&      3.5&3.0&2.8& 0.5&0.4&0.4&0.3&0.053(0.17)\\
Ib47&100K&  83.9&69.2&   26.7&    25.5&21.5&  20.2& 3.6&3.1&3.0&2.2&0.39(1.23)\\
Ib57&100K&  13.7&11.2&  4.3&       4.1&3.4&3.2& 0.6&0.5&0.5&0.4&0.07(0.22)\\
\hline
IIa11&100K&8.5& 7.0& 2.7&   2.5&    2.1&2.0& 0.3& 0.3 & 0.3&0.2&0.035(0.11)\\
IIa14&100K& 4.9&4.1& 1.6&   1.5&    1.3&1.2&  0.3&0.17& 0.16&0.12&0.021(0.07)\\
IIa26&100K& 3.8&   3.1&     1.2&    1.1&0.9&  0.9& 0.1& 0.1&0.1& 0.1&0.02(0.06)\\
\hline
Ya11&100K& 5.4&   4.4&  1.7& 1.6&1.4&         1.3&0.2&0.2& 0.2&0.1&0.02(0.06)\\
Ya12&100K& 4.8&   4.0&  1.5& 1.4&1.2&         1.1& 0.2& 0.2& 0.2&0.1&0.02(0.06)\\
Ya14&100K& 4.9&  4.0&   1.6& 1.5&         1.2& 1.1& 0.2& 0.2&0.1& 0.1&0.02(0.06)\\
\hline
\hline
$\nu t \bar b$&100K& 50712.1& 28338.4& 15293.7& 12381.7& 10299.7& 9465.2& 209.7& 144.5& 75.9& 13.2&\\
$\nu b\bar bj$&560K& 14104.6& 6122.8& 3656.7& 2568.2& 2465.8& 2272.4& 103.7& 60.8& 37.4& 8.7&\\
$\nu b2j$&90K& 18043.1& 8389.2& 3013.0&2744.8& 2278.1& 2171.4& 99.5& 40.0& 25.2& 5.3&$B$=31.7\\
$\nu 3j$&300K& 948064.2&410393.4& 15560.9& 0.0& 0.0& 0.0& 0.0& 0.0& 0.0& 0.0&$\sqrt{B}$=5.6\\
$e b \bar b j$&115K& 256730.1&55099.8& 36353.6& 25010.7& 2638.4& 453.3& 29.3& 18.0& 11.3& 4.5&\\
$e t\bar t$&130K& 783.3& 685.0&384.5& 298.8& 204.5& 29.9& 1.0& 0.8& 0.5& 0.0&\\
\hline
\end{tabular}
}
\caption{Same as Tab. \ref{tab:H130} but for $m_H$=170 GeV.}
\label{tab:H170}
\end{table}

The analysis strategy has been adopted from earlier work of some of us \cite{Das:2010ds}. In particular,
we have exploited a simple cut-based method for signal enhancement and 
background rejection. We have chosen the following selections
and applied them cumulatively for the signal from $h$ ($H$).

\begin{figure}[ht]
\centering
 \includegraphics[width=2.9in]{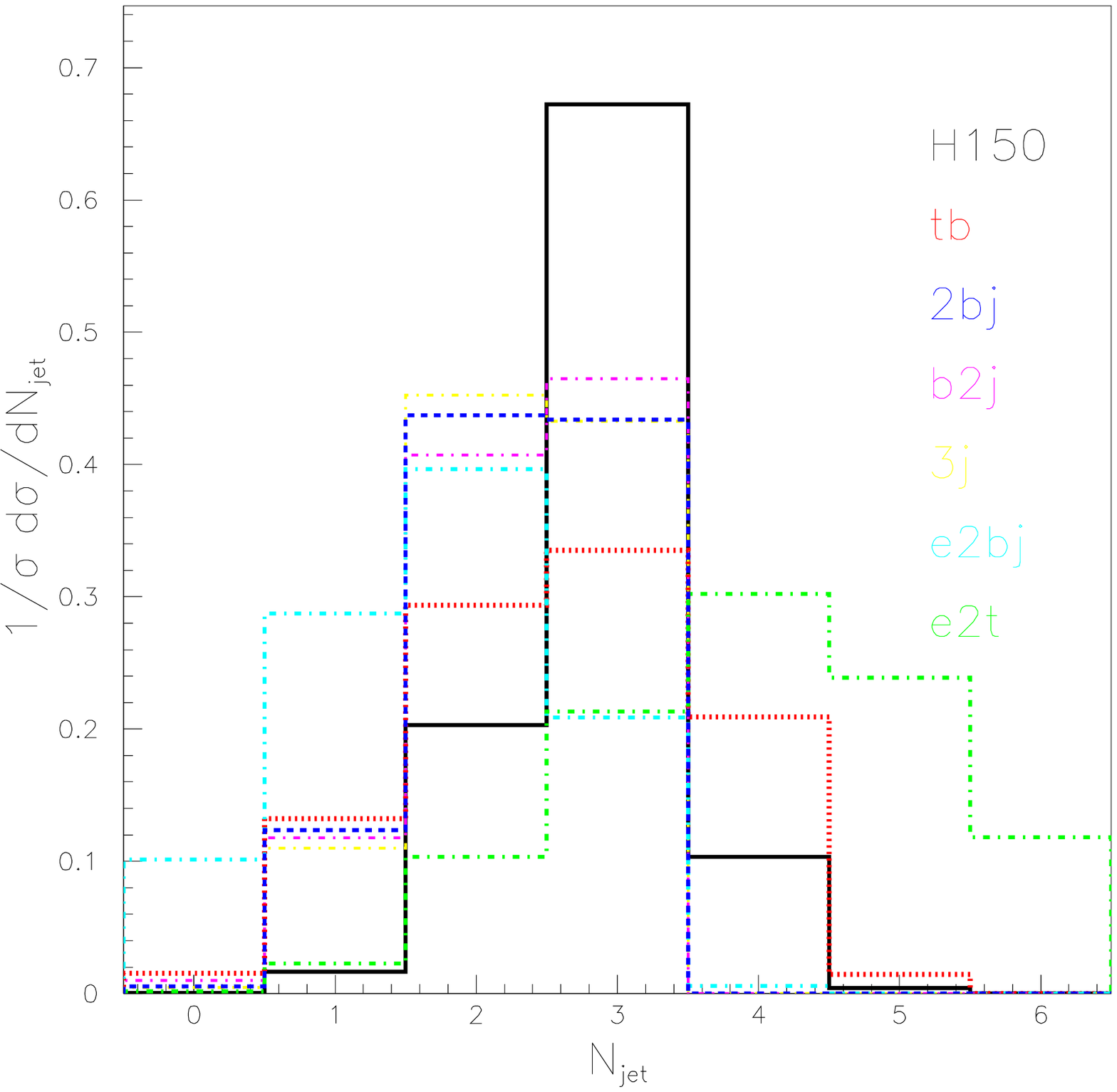}
 \includegraphics[width=2.9in]{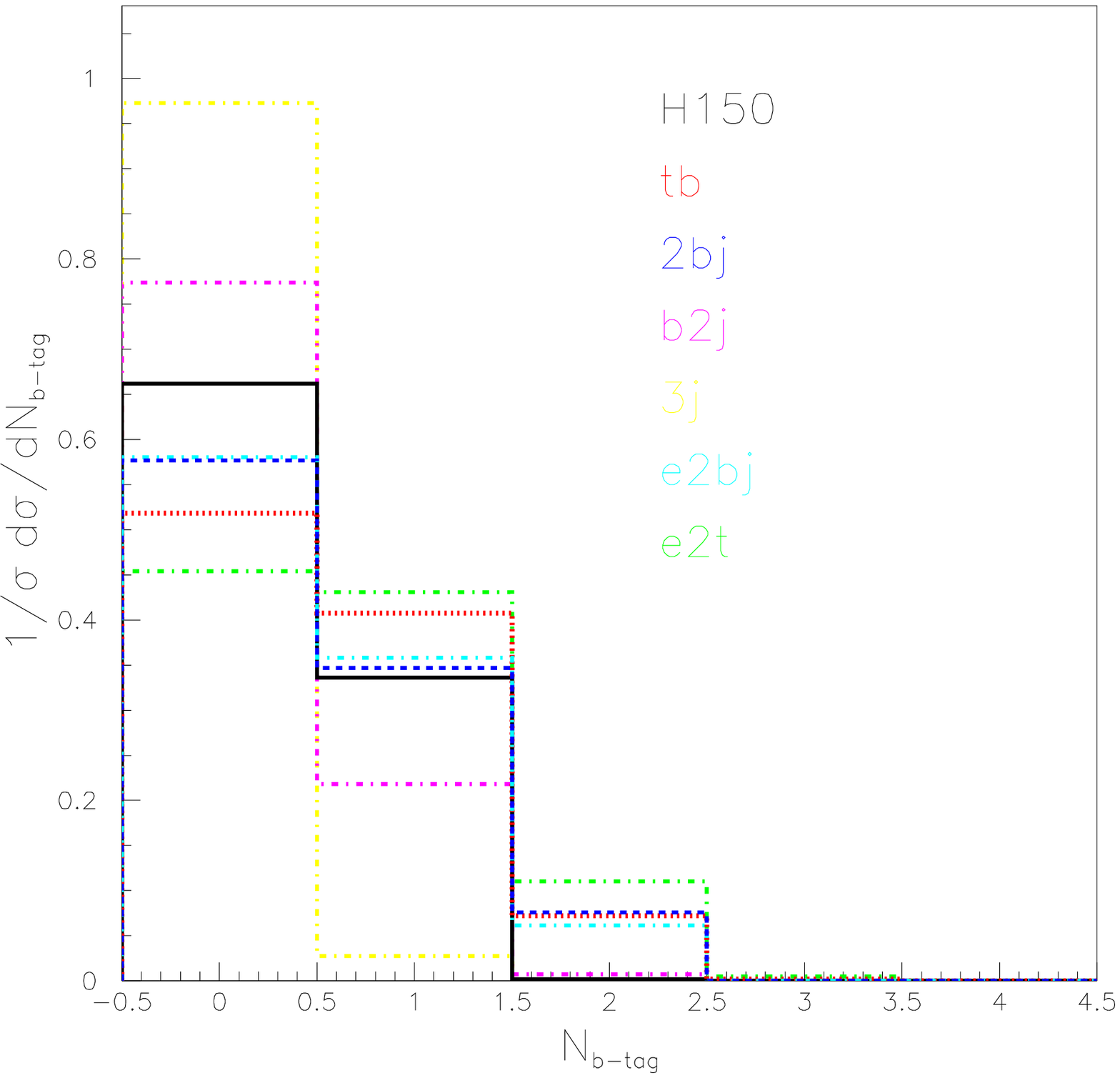}
\caption{\label{njetnbtag} Scenario Ib with the parameter $X=28=Z$ and $Y=10$. The number of jets ($N_{jet}$) in the left panel and the number 
of $b$-tagged jet with the inclusion of mis-tagging ($N_{b-tag}$) in the right panel 
for signal ($m_H$=150 GeV) and all the SM backgrounds. For other signal 
events, the distribution profiles are very similar, except the fact that 
the number of jets as well as $b$-tagged jets is slightly  larger for 
heavy Higgs bosons. See the fourth(fifth) column for their actual efficiencies 
with selections applied in a(b) and A(B) for lighter(heavier) 
Higgs bosons, respectively.}
\end{figure}

\begin{itemize}

\item {\bf a(A):} We first selected events containing at 
least three jets (same). 
The distribution of the number of jet ($N_{jet}$) is shown in the 
left panel of Fig. \ref{njetnbtag}.
For the lighter Higgs, $h$, in all the signal benchmarks the efficiencies\footnote{Unless 
mentioned otherwise, all the efficiencies quoted hereafter are given with respect to the 
previous selection.} are approximately 70\%(For heavier Higgs boson $H$, with mass of 130, 150 and 170 GeV the
efficiencies are 71\%, 76\% 81\% respectively, as two of the jets are directly 
coming from the corresponding heavier Higgs boson and hence the efficiencies are larger.). Further, $t\bar b$ has efficiency of
approximately 56\%, $2bj$ and $3j$ events display approximately 45\% 
whereas the $b2j$  one has approximately 47\%. Efficiencies are generally lower for 
the photo-production channels: $e2bj$ is approximately 20\% (the sharp fall is due to the 
isolation criterion) whereas for $e2t$ 
the jet efficiencies are higher and due to the presence of two $t$-quarks 
leads to two $b$-quarks and the probability of having two jets 
from $W$-boson decay itself is approximately 91\%, thus, out of 
4-jets in 91\% events, the probability of having at least three 
energetic jets is reduced by 4\%, which leads to approximately 
87\% efficiency. However, as we will see, the presence of 
the electron in the photo-production modes leads to the corresponding backgrounds
to be very low.

\item {\bf b(B):} We demanded at least one $b$-tagged jet 
with the inclusion of proper mis-tagging (same).
The distributions of the number of $b$-tagged jets ($N_{b-tag}$) 
are shown in the right panel of Fig. \ref{njetnbtag}.
For the lighter Higgs, $h$, in all the signal benchmarks the $b$-tagging 
efficiencies are approximately 40\%.  
In fact, all our signal benchmarks contain at least one $b$-quark and, since we adopted $\eps_b$=0.50,
the 10\% lowering is quite realistic and due to the fact that not 
all $b$-quarks in the signal are eligible for the $b$-taggable criterion adopted in our 
analysis. For heavier Higgs signals, benchmarks show similar 
efficiencies and changes are less than 1\% for three masses. In case of $t \bar b$, the events containing 
at least one  $b$-tagged jet are approximately 54\%, a rate greater than the 
signal ones, as these background topologies contain at least two $b$-quarks at the parton 
level so that, due to combinatorics (other than mis-tagging a
light-flavor quark-jet from $W$-boson decays), the probability of 
one $b$-tag is more. The probability of $2bj$ is approximately 60\%, 
approximately 6\% larger than $t \bar b$, due to,  
unlike $t \bar b$, the presence of one $b$-quark and one light-flavor jet 
 in the hard processes. Further, $b2j$ efficiencies are
similar to or little less than those of the signals due to the fact that 
the taggable rapidity is more central, where the jets are more likely to be
forward in the basic hard processes. The efficiency of 
$e2t$ is approximately 2\% larger than for $t\bar b$ mainly due to the contributions 
from mis-tagging one extra hadronic $W$, i.e, two extra jets. Finally, 
the efficiency of $3j$ is approximately 4\%, which seems  consistent 
with the expected mis-tagging rates (1.0\% for light-flavor jets and 
10\% for $c$-jets) in presence of combinatorics.

\item {\bf c(C):} We demanded at least two central jets, with $p_T$ $>$ 30 GeV 
($p_T>$ 25, 20 and  15 GeV for $m_H$=130, 150 and 170 GeV, respectively\footnote{Since the 
cross sections become smaller with increasing $m_H$, we lowered 
the central jet $p_T$ cuts.}) in the pseudo-rapidity range $|\eta| <$ 2.5. One of 
the central jet must be a $b$-tagged jet and we demanded only one $b$-tagged 
jet (same).
For the lighter Higgs, $h$, all the signal benchmarks as well as the $t \bar b$ noise 
survive at the rate of approximately 65\%  since all these processes 
naturally have three jet in their events. Further, $e2bj$ and $2bj$ are reduced by 
approximately 35\% and 50\%, respectively, mainly due to demanding, 
with respect to the b(B) case above, of one $b$-tagged jet only.
The diagrams of $tb$ and $e2t$ reveal that these two backgrounds are more central,
because of the presence of one $t$-quark in the central region. In fact, the 
efficiency is larger in $e2t$ and is mainly due to the contributions from the additional 
$t$-quark. Although in $2bj$ the probability in presence of one $b$-tagged in 
the central region is large  the overall efficiency is reduced to 12\% 
due to (partly) the possibility of more than one $b$-tagged jets whereas 
for $Wb$ the value is mainly due to the hadronic branching fraction
and also that the $b$-tagged jet is not necessary central. 
In case of $e2bj$ the efficiency is 35\%: this noise suffers mainly due
to the centrality criterion. For $3j$, none of the events survived 
this selection criterion.
The efficiencies pattern discussed above are similar for the heavier Higgs boson, $H$, 
yet recall that here we have used slightly softer selections
on the transverse momentum. Thus, the efficiencies are increasing with a softer $p_T$ 
selection for both signals and backgrounds.

\item {\bf d(D):} The missing transverse energy cut $\met$ $>$ 20 GeV is first applied (same).
The relevant distribution is shown in the left panel of Fig. \ref{metetaf}.
\begin{figure}[ht!]
\centering
 \includegraphics[width=2.9in]{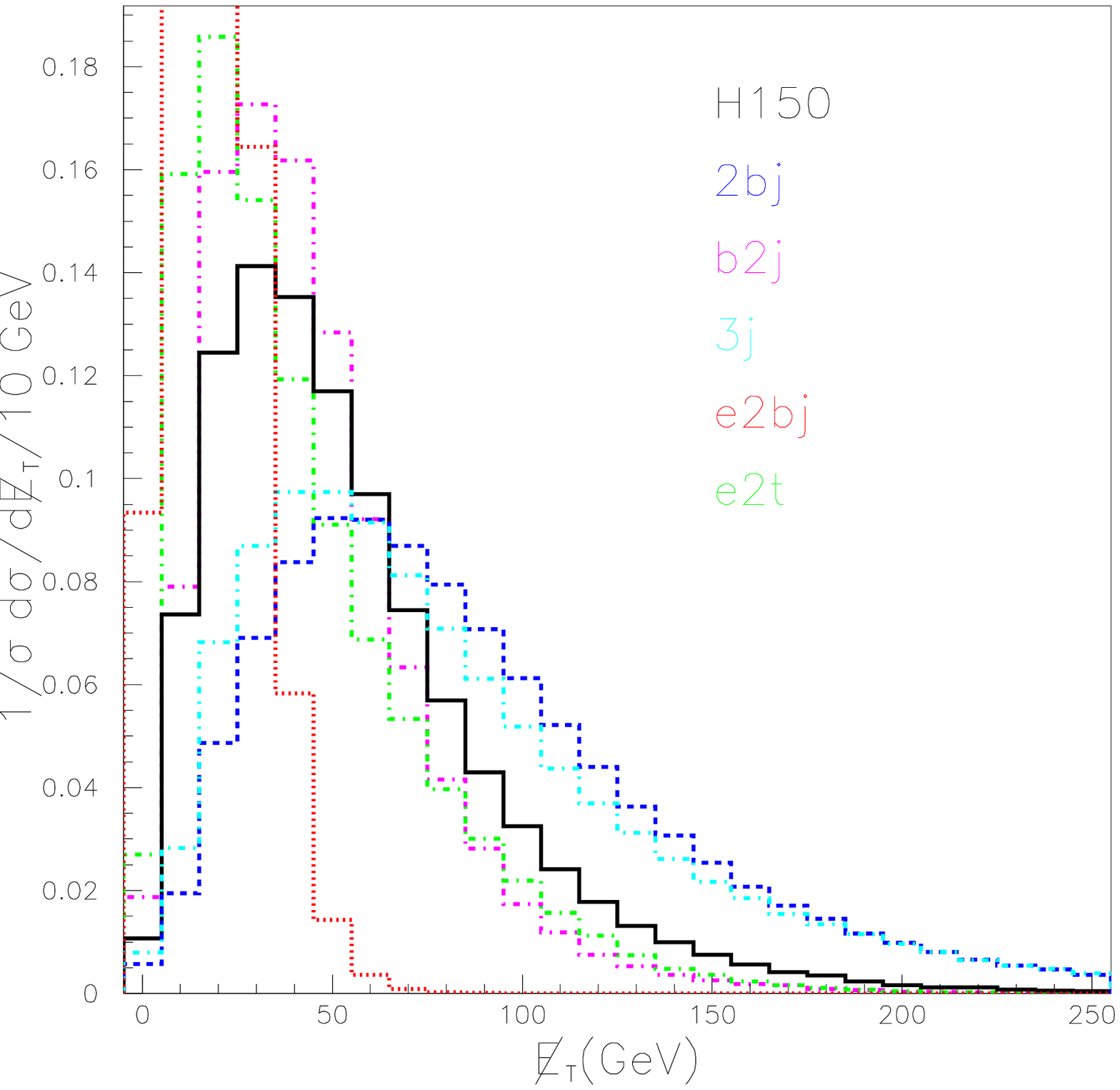}
 \includegraphics[width=2.9in]{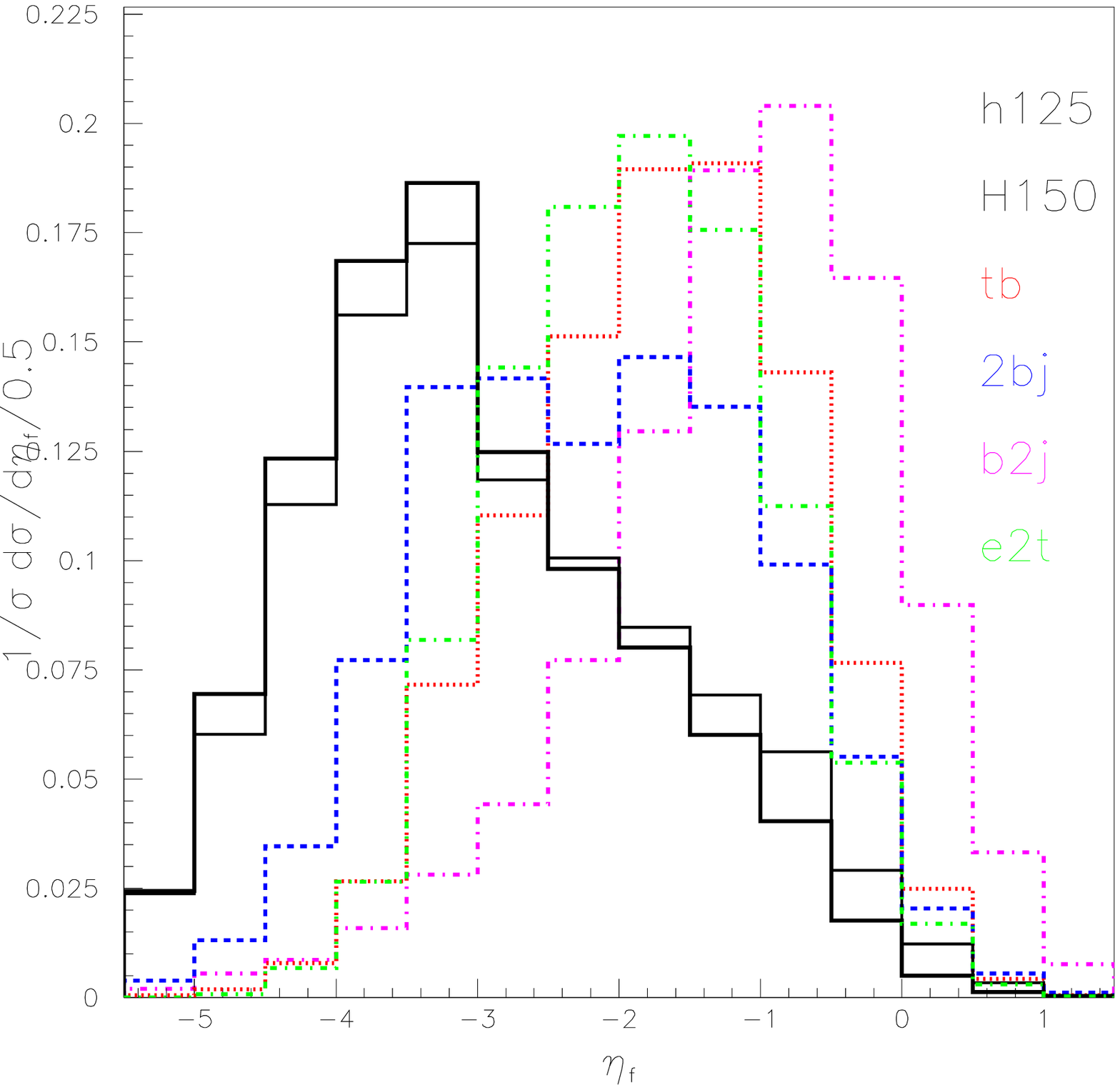}
\caption{The missing energy ($\met$) (left panel) and  rapidity 
($\eta_{j_f}$) (right panel) profile of the forward jet for signals and SM backgrounds. The $\met$ distributions 
for all other signal benchmarks as well as the $t \bar b$ noise are not shown as they are very similar 
to the signal distributions of $m_H$=150 GeV for Scenario Ib with $ X=Z=28$ and $Y=10$ (shown in thick solid),
whereas the thin solid is for $m_h$=125 GeV for Scenario Ia with $X=Z=28$ with $Y=10$. The rapidity 
distributions profile for $m_H$=130(170) GeV is very close to the $m_h$=125 GeV($m_H$=130 GeV) case
shown in thin solid, except  that for massive Higgs the peaks shift towards the left.
Also the corresponding rapidity distribution profile for {\it e2bj} is somewhat similar to the $m_h$=125 GeV signal case. }
\label{metetaf}
\end{figure}
For all the signal benchmarks (lighter as well heavier Higgs bosons),  
$t \bar b$ and $b2j$ the efficiencies are approximately 83\%. The sources of neutrinos 
and the event structures of these two processes are very similar, 
except for the fact that the top-quark decays produce either 
neutrinos (which are then pure sources of missing energy but in such a case, owing to the selection a(A), 
they are largely removed) or quarks (where the
smearing of jets and track mis-measurements are the main sources of missing energy). For $2bj$ the 
efficiencies are approximately 96\%. This selection is crucial to suppress 
the photo-production processes: $e2bj$ and $e2t$. In case of $e2bj$ only 
12\% of the events survive in fact. For $e2t$ the presence of two $W$-bosons and their decays 
into leptonic modes would be the sources of missing energy, so that this noise is not affected very much
by our constraints.

\item {\bf e(E):} A lepton ($e$ or $\mu$) veto for $p_T$ $>$ 20 GeV and 
$\eta$ $<$ 3.0 is applied here (same). 
For the lighter Higgs signal benchmarks, the efficiency for this 
selection is approximately 96\%, as only 4\% of the  events contain at least 
one lepton which is coming from the semileptonic decays of  
$B$-hadrons and $D$-mesons and which passes the isolation criterion above. 
The efficiencies for $2bj$ and $b2j$ are rather close to the 
signal benchmarks, due to these processes also not having prompt leptons 
in their events. The $t\bar b$ channel  has an  efficiency of 93\%, hence
approximately 3\% less than the signal, as here the top-quark decays 
can lead to one bottom-quark and, if the hard-processes bottom quarks are 
more central, the  requirement in c(C) is satisfied and there 
is no problem in having a central and high-$p_T$ lepton from a $W$-boson decay. In the photo-production 
processes, $e2bj$ and $e2t$ both contain hard leptons, so only 15\%  of the events 
survive our  lepton veto.
For heavier Higgs masses, the efficiencies are quite similar 
 to the lighter Higgs boson mass case. The only difference
is that, since the applied $p_T$ threshold is lower for heavier Higgs masses,
the probability of having a lepton in the event is higher, thus 
the veto efficiencies are somewhat smaller, except for the two photo-production 
processes.

\item {\bf f(F):} In the central region, defined above via c(C), we reconstruct 
the invariant mass of one $b$-tagged jet with any of the other jets, $M_{bj}$. Amongst these,
we have chosen the best combination, i.e., where, the absolute difference 
$|M_{bj} - M_{h(H)}|$ is minimized. We call this di-jet combination the candidate 
light (heavy) Higgs boson signal. In order to select the latter, we have kept events 
within a 15 GeV mass window centered around the corresponding Higgs boson masses. 
The distributions of $M_{bj}$ are shown in the left panel of Fig. \ref{mbsmbsj}.
It seems that the di-jet invariant masses of the {BP-Ia30} signal 
benchmark (with $m_h$=125 GeV) has its peak around 115 GeV\footnote{The peaks 
always show up to the left side of the actual masses due to jet energy smearing 
and the shift also depends on the jet-cone size.}. The distribution for $t\bar b$ also has a 
peak around that of the $m_h$=125 GeV signal. However, the combinatorics is significant and this shows in 
their  efficiencies, which are approximately 40\% for both. 
The distribution of $b2j$ is flat as there is no correlations for the 
correct di-jet candidates. Also note that $b2j$ has $W$-boson exchange resonant 
diagrams, so a probability in principle exists for a di-jet invariant mass
peak at $M_W$, however, this  is very small, mainly due to low 
mis-tagging efficiencies and the centrality criterion. Further, also in case of 
$2bj$, where the $Z$-boson is present resonantly in the diagrams, the $Z\to b\bar b$ 
decay combined with high tagging efficiencies allows for the appearance of a di-jet peak at 
80 GeV (approximately 10 GeV less than $M_Z$ due to jet energy smearing): 
see the left-panel of Fig. \ref{mbsmbsj}.
In case of $e2t$, like $t \bar b$, one has also correlated di-jet candidates, 
but the energy scale is higher, so the peak is shifted to higher masses. The 
efficiency is approximately 45\%, a little larger than the signal 
and $t\bar b$ ones. The distributions of $b2j$ and $e2bj$ are flat and the 
efficiencies are the same, approximately 28\%. This particular selection 
suppresses $2bj$ events more severely though, at a rate which is approximately 15\%.
For heavier Higgs bosons the distributions show rapid falls and 
so, by applying the mass window cuts, only the left part of the 
distributions contributes. This shows in their signal efficiencies, 
which are approximately 34\%, 23\% and 18\% for $m_H$=130, 150 and 
170 GeV, respectively.  
The SM backgrounds do not show up in distributions at large invariant masses,
thus for heavier Higgs mass combined with the same mass window selection suppresses
more the backgrounds. As an example, in case of $e2t$, which produces somewhat 
higher invariant masses than all other SM backgrounds, the efficiencies 
drops to 40\%, 12\% and 4\% for $m_H$=130, 150 and 170 GeV, respectively. 
In case of $t \bar b $ the efficiencies (see 
Tabs. \ref{tab:h125}, \ref{tab:H130}, \ref{tab:H150} and \ref{tab:H170}) 
drop to from 30\%, 8\% and 2\%, respectively.
In case of $2bj$ the values are 14\%, 10\% and 5\%. For 
$e2bj$, one has 20\%, 10\% and 7\%, respectively. Finally,
for $b2j$, these are 25\%, 12\% and 5\%, respectively.

\begin{figure}[ht!]
\centering
 \includegraphics[width=2.9in]{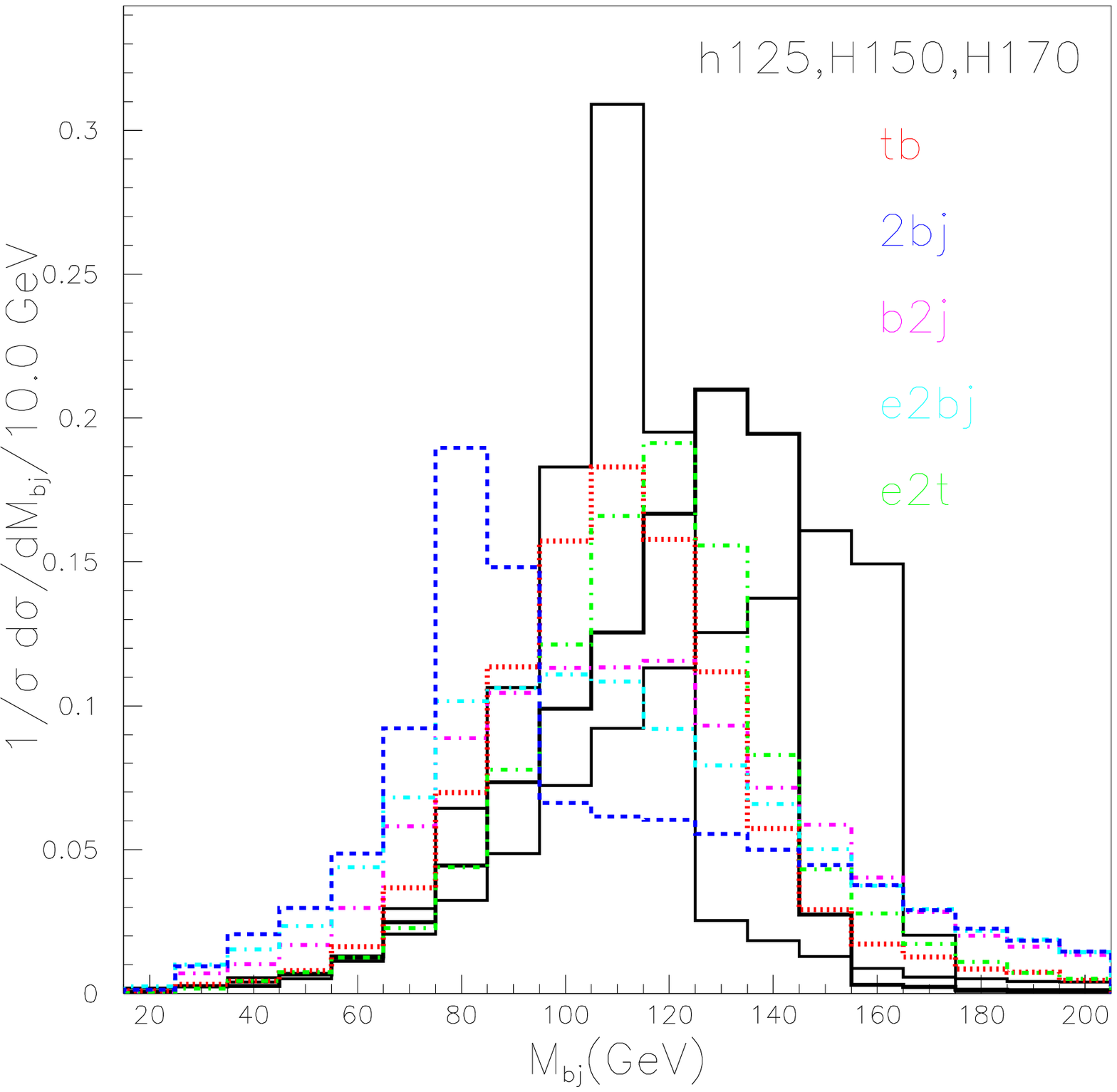}
  \includegraphics[width=2.9in]{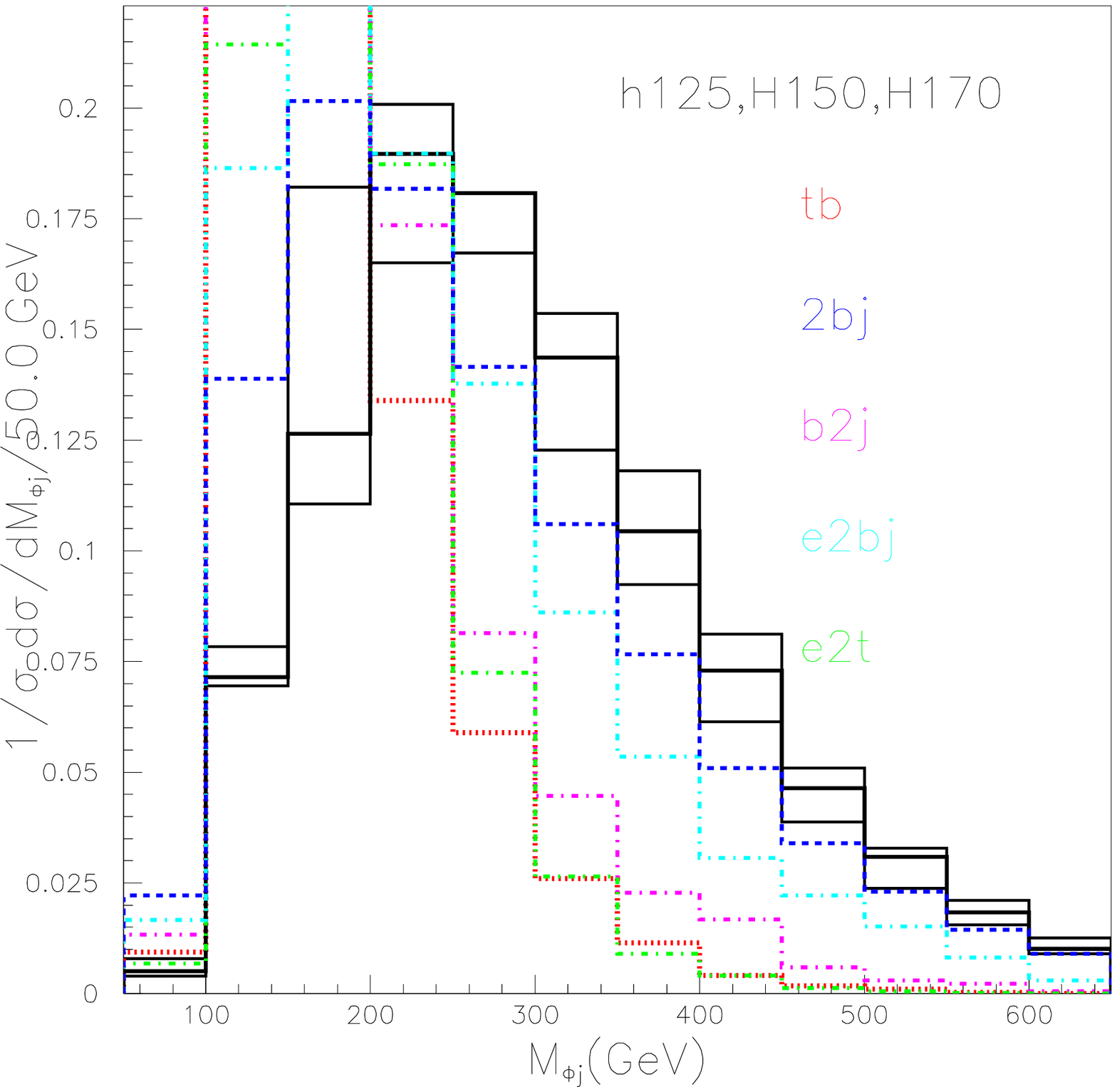}
\caption{The di-jet invariant mass, made up by  one $b$-tagged and one light-flavor jet, producing Higgs 
candidates, $M_{\phi}= M_{bj}$ (left panel) and the three-jet invariant mass, i.e., the previous two jets combined together with the 
forward jet, $M_{\phi j_f}$ (right panel). The mass peaks of the Higgs signals ($M_{\phi}$) 
correspond to $m_h$=125 (thin black) for Scenario Ia, $m_H$=150 (thick black)  and 170 (thin black) for Scenario Ib from left to right. All these 
are using the parameters $X=Z=28$ and  $Y=10$.
The distribution for $m_H$=130 is not shown but it lies in between $m_h$=125 and $m_H$=150. 
Among all SM backgrounds, only $2bj$ shows a prominent peak from the $Z$-boson. Notice that  
$M_{\phi j_f}$ represents the overall energy scale of the hard-scattering.
}
\label{mbsmbsj}
\end{figure}

\item {\bf g(G):} We demanded the remaining leading jet in the event to have  $p_T$ $>$ 25 GeV, 
with $-5.5 < \eta < -0.5~(-1.0)$  (these values are chosen by seeing the distribution, 
see the right panel of Fig. \ref{metetaf}) and termed it as the forward  jet ($j_f$).
This forward jet lies very close to the direction of the incoming proton, like in
Vector Boson Fusion (VBF) processes for Higgs production. In contrast to VBF though, 
instead of a jet with large rapidity gap with respect the the forward jet (a backward jet then), in our signal we have a neutrino. 
The more massive the Higgs is, the less energy remains for the forward jet so as to lay close to the proton direction, i.e., at larger rapidity. This reflects in the 
right panel of Fig. \ref{metetaf}. The thick (thin) solid curve corresponds to 
$m_H$=150 GeV ($m_h$=125 GeV).
For a lighter Higgs boson, $m_h$=125 GeV, the efficiency is approximately 80\%.
For $e2t$ the efficiency is almost 90\%, twice that of $e2bj$, as there is 
more than twice a probability to have a forward-jet from top-quark decays.

\item {\bf h(H):} The di-jet invariant mass of the Higgs boson candidates 
with the forward tagged jet, which is essentially  the overall energy scale 
of the hard scattering, is asked to comply with the following requirements: 
$m_{h j_f}$ ($m_{H j_f}$) $>$ 190 GeV (190, 
210 and 230 GeV for $m_H$=130, 150 and 170 GeV, respectively).
The distributions are shown in the right panel of Fig. \ref{mbsmbsj}.
For a lighter Higgs boson, $m_h$=125 GeV, except a few cases\footnote{Recall that 
the efficiencies are relative to the previous selection, one can estimate the individual efficiencies from the 
respective distributions.}, the efficiency is approximately 82\%. 
This forward jet should not be a $b$-jet though. So, in $t\bar b$ and $e2t$, where a 
forward $b$-tag jet is more probable, the efficiencies are lower, approximately 
 32\% and 37\%.
It is clear from the right panel of Fig. \ref{mbsmbsj} that the three-jet invariant mass 
distributions of $b2j$ peak around 140 GeV or so. The same for $t \bar b$, $e2t$, $2bj$,
$e2bj$ and the Higgs signal with $m_h$=125 GeV, which show somewhere around 180 GeV. So, 
for $m_h$=125 GeV, the efficiency is around 82\%.
For the heavier Higgs boson, with $m_H$ = 150 (170) GeV, the distributions are shown 
in the thick (thin) solid  curve in the right panel of Fig. \ref{mbsmbsj}
and peak around 220 (260) GeV. The selection cuts for these two Higgs bosons 
are 210 and 230 GeV, respectively. With our selection, for these two 
heavy Higgs signals, one suppresses more SM backgrounds than in the case of the Higgs 
signal with $m_h$=125 GeV and $m_H$ = 130 GeV.  For example, 
in the $t \bar b$ case, the most dominant background, for $m_h$=125 GeV, 
and $m_H$=130, 150 and170 GeV, the events which survived are approximately 
652, 618, 195 and 80, respectively. For other SM backgrounds a similar 
pattern follows, thus the overall SM backgrounds can be strongly reduced.
However, this  overall background rejection will not help alone to have  
larger significances as the signal rate itself is suppressed with heavier 
Higgs masses (at the production level).

\item {\bf i(I):} Finally, we required only one light-flavor jet in 
the central region (same).  This selection is called `central jet veto' and has severe impact on 
all  processes having more jets in the central rapidity region, other than 
the Higgs candidate jets. Recall that our Higgs signal candidate 
jets, selected in f(F) above, are central: this is true for
not only the signal, but also the dominant SM background, $t \bar b$.
For a lighter SM Higgs, see Tab. \ref{tab:h125}, approximately 35--40\% of the events 
have a central jet other than Higgs candidate jets, thus only 60--65\% of the events
survive. For $t \bar b$, $\nu b 2j$ and $\nu 2b j$ the efficiencies 
are 22\%, 18\% and 14\%, respectively. Among all the SM backgrounds, $e2t$ has a larger 
number of jets (see the distributions in the left panel of Fig. \ref{njetnbtag}),
thus the probability of having a central jet is more, so that this selection 
suppresses this background severely, approximately by 93\% (for all the 
Higgs cases, see Tabs. \ref{tab:h125}, \ref{tab:H130}, \ref{tab:H150} and \ref{tab:H170}). 

\end{itemize}

After the cumulative selections from {a--i}, discussed above, we find that, 
for the SM Higgs boson with $m_h$=125, the final number of events is around  15--30 only 
for Scenario  Ib and for large values of the parameters $X=Z=28 (44)$ and $Y=10 (5)$ respectively. The total 
SM background rate is approximately 170. The charged-current backgrounds, 
$\nu t \bar b$, $\nu b \bar b j$ and $\nu b 2j$, are the dominant ones and 
only 3\% of the total background comes from $e t \bar t$ photo-production.
These rates lead to a maximum significance of approximately 2.4 (7.5)$\sigma$  with 100  (1000)$fb^{-1}$ integrated luminosity for 
Scenario Ib with $X=Z=44$ and $Y=5$. For Scenario Ib with $X=Z=28$ and $Y=10$, the significance is approximately
1.2 (3.8)$\sigma$ for 100  (1000)$fb^{-1}$ integrated luminosity. The significances for Scenario Y and Scenario IIa are less than 1.
Thus, one can expect that Scenario Ib with large value of $X=44=Z$ and $Y=5$
may be accessible through the SM-like Higgs boson signal already detected at the LHC.

We also searched for the second CP-even neutral Higgs boson of our 2HDM-III, with masses $m_H$=130, 150 and 170 GeV.
After the cumulative selections from {A--I},
the maximum number of signal events for $m_H$=130 GeV is approximately 15 (30) and only 
for Scenario Ib with $X=Z$=44 (30) with $Y=5$. The total SM background is approximately 150.
So the maximum significance is approximately 2.6 (8.1)$\sigma$  for 100  (1000)$fb^{-1}$ integrated luminosity. For the case $m_H$=150 GeV, 
the number of signal events is approximately 7 and the SM background 
reduces to approximately 60: this leads to a significance of approximately 1.0$ (2.6)\sigma$ with 100  (1000)$fb^{-1}$ integrated luminosity.
For $m_H$=170 GeV, the raw event rate is approximately 80 to start with and, at the 
end, the count is only 2\footnote{Note that our selection cuts applied above are not optimized. An increase 
of the luminosity is an easy solution from a phenomenological perspective. However, 
adopting multivariate analysis techniques must also be a better discriminator 
of signal from backgrounds.}. The total SM backgrounds is approximately 30, which 
leads to a significance approximately 0.4 (1.23)$\sigma$. 

The LHeC will be operational for about ten years and expected to accumulate a
total integrated luminosity of 1000 fb$^{-1}$ of data. So, at the end of the run, we expect 
the SM Higgs boson will have 7.5$\sigma$ (3.8$\sigma$) significance for Scenario Ib 
with $X=Z$=44 (28) with $Y=5 (10)$ . For Scenario IIa and 
Scenario Y, with $X$=26 and $Y=2$, the final significances are approximately 1$\sigma$. It seems that in all scenarios of the model, large $X$ are favorable.
For heavy Higgs masses with $m_H$=130 GeV,  for Scenario Ib with $X$=44 and 30 with $Y=5$, 
the maximal significances are approximately 8.1 and 3.7$\sigma$ respectively. For $m_H$=150 GeV,  Scenario Ib 
with $X$=44 and $Y=5$, the final significance is 2.62$\sigma$. For $m_H$=170 GeV, 
in the Scenario Ib with $X$=30 and $Y=5$, the final number of signal events is 
approximately 2. The estimated significances is 1.23$\sigma$. Thus, for high enough 
Higgs masses, one might invoke the aforementioned multivariate analyses to have larger significances.

\section{Conclusions}
\label{sec:conclude}

After the discovery  of a SM-like Higgs boson at LHC, one is well motivated to look 
for more such states, which necessarily appear in BSM scenarios. Among the experimental facilities where more Higgs bosons can be searched for, one should list 
an $ep$ collider which may be built at CERN, known 
as the LHeC. In our analysis we have considered a 2HDM-III with a four-zero Yukawa texture
 in three configurations, wherein both the SM-like Higgs boson and the heavier version of it
can be accessible at the foreseen LHeC energy. We assumed that both of 
these states are decaying via a flavor-violating mode ($\phi \to b \bar s$). 
After a parameter scan, 
we have selected a few model benchmarks where the products of cross sections and  
flavor-violating BRs are large enough to produce sufficient 
events in which to look for both signatures.
We studied the three-jet and missing energy channel, $3j + \met$, from the 
charged-current production of $\nu_e \phi q_f$, where $q_f$ is a forward 
jet with large rapidity and the other two jets come from the flavor-violating 
decay $\phi \to b \bar s$. We demanded one central jet to be $b$-tagged 
with the inclusion of the proper mis-tagging from light-flavor and gluon jets. 
We considered the most dominant SM backgrounds: charged-currents, 
$\nu t \bar b$, $\nu b \bar b j$, $\nu b2j$ and $\nu 3j$, and 
photo-production, $e^{-} b \bar b j$ and $e^{-} t \bar t$.
We performed a full hadron--level Monte Carlo simulation using {\tt CalcHEP} as matrix element calculator,  
{\tt PYTHIA} as parton shower/hadronization event generator and its {\tt PYCELL} 
toy calorimeter in accordance with the LHeC detector parameters. 
We carefully implemented $b$-tagging, including mis-tagging of 
$c$-jets or light-flavor or gluon jets. 

The signals under consideration do not have leptons, so we applied lepton vetos. The 
charged-current production has naturally missing energy due to the presence of 
neutrinos but no charged lepton. However, the photo-production processes have leptons in them 
but no direct missing energy (except the mis-measurements from  jets and smearing), 
thus the missing energy selection together with the lepton veto suppressed the 
photo-production backgrounds to a very large extent.

The kinematics of the particular signals considered is very interesting 
from the fact that the Higgs boson is produced in the central rapidity region 
and its decay daughters, one $b$-jet and one light-flavor jet, are also central. 
We reconstructed the invariant mass of this two jets and selected events
only for  masses within a 15 GeV window around the respective Higgs masses
of the signal benchmarks. This selection reduces the SM backgrounds to a large 
extent and the invariant mass ensures the selection of  flavor-violating decays. 
For massive Higgs bosons, although the signal events becomes low with 
the mass window selection, background suppression is more efficient. 

As a next step of our selection, we identified the most energetic light-flavor 
forward jet (by seeing the rapidity profiles) and calculated the invariant 
mass with that jet together with the flavor-violating Higgs candidates jets. 
These three-jet invariant masses essentially give the overall energy scale 
of the hard scattering. Again, the more massive  Higgs boson   
helps to suppress more SM backgrounds, in particular $\nu t \bar b$ and $e t \bar t$,
but the signal becomes smaller too.

At the end, the most important  cut, we applied a central jet veto, i.e., 
 we required one light-flavor central jet only. This suppresses SM 
backgrounds with large multiplicity, for example, $\nu t \bar b$ and $e t \bar t$.

After all the selections, with 100 fb$^{-1}$ of data, we found that the SM Higgs 
boson, $h$, would  be detectable within the 2HDM-III in the scenario called in this work  Ib with$X=Z=44$ or 30 with $Y=5$, with approximately 1-2$\sigma$. The heavier neutral Higgs boson, $H$, 
with masses 130 GeV (150 GeV), would have  2 (1)$\sigma$ significances for large $X$ and only for Scenario Ib. 

The LHeC will be operational for around ten years and so it is expected to accumulate a
total integrated luminosity of 1000 fb$^{-1}$ of data. So, in all the cases 
mentioned above, the final significances will be enhanced. At the 
end of the run, the 2HDM-III Like-IIa (Like-Y) the SM Higgs will have 
1$\sigma$. For $m_H$=130 GeV, in the Scenario Ib with 
$X$=44  (30) and $Y=5$, the maximal significances are approximately 3.7 (8.1)$\sigma$. 
The maximal significances for $m_H$=150 GeV is 2.6$\sigma$ for Ib with $X$=44 and $Y=5$.
For $m_H$=170 GeV the final number of signal events is approximately 2, probably too little to be detected. 
However, it should be noted that 
we have adopted a simple cut-based method in this analysis. One would instead invoke more complex 
discriminators to enhance the significances within the designed luminosity, for example, 
multi-variate analyses.

To conclude, after the first few years of the LHeC running, by adopting more complex discriminator
and/or multi-variate analyses, we expect that both $h$ and $H$ signals will appear at the LHeC. 

\acknowledgments

This work has been supported by SNI-CONACYT (M\'exico), VIEP-BUAP and by 
PRODEP-SEP (M\'exico) under the grant: ``Red Tem\'atica: F\'{\i}sica del Higgs y del Sabor". 
RX acknowledges the scholarship from CONACYT (M\'exico).
S.M. is supported in part through the NExT Institute. SPD acknowledges travel grant 
and all other supports from PRODEP-SEP (M\'exico): 
``Red Tem\'atica: F\'{\i}sica del Higgs y del Sabor". SPD also acknowledges the 
visiting fellowship from FCFM and support from 
the project ``New Physics with CMS''. SPD is grateful to the post-doctoral 
fellowship and academic leave from Institute of Physics, Bhubaneswar, Odisha, India, 
while the project started.

\end{document}